\documentclass[preprint2]{aastex}
\usepackage{color}
\usepackage{epsfig}
\usepackage[fleqn]{amsmath}
\usepackage{multirow}
\usepackage{mathptmx}
\usepackage{graphicx}
\usepackage{bm}

\shorttitle{Explicit symplectic integrators} \shortauthors{Wu et
al.}

\begin{document}


\title{Explicit symplectic integrators with adaptive
time steps in curved spacetimes}


\author{Xin Wu$^{1,*}$, Ying Wang$^{1}$, Wei Sun$^{1}$, Fuyao Liu$^{1}$, Dazhu Ma$^{2}$}
\affil{1. Center of Application and Research of Computational
Physics $\&$ School of Mathematics, Physics and Statistics,
Shanghai University of Engineering Science, Shanghai 201620, China
\\ 2. College of Intelligent Systems Science
and Engineering, Hubei Minzu University, Enshi 445000, China }
\email{Emails:  $*$wuxin$\_$1134@sina.com (X. W.)} 


\begin{abstract}

Recently, our group developed explicit symplectic methods for
curved spacetimes that are not split into several explicitly
integrable parts, but are via appropriate time transformations.
Such time-transformed explicit symplectic integrators should have
employed adaptive time steps in principle, but they are  often
difficult in practical implementations. In fact, they work well if
time transformation functions cause the time-transformed
Hamiltonians to have the desired splits and approach 1 or
constants for sufficiently large distances. However, they do not
satisfy the requirement of step-size selections in this case.
Based on the step-size control technique proposed by Preto $\&$
Saha, the nonadaptive time step time-transformed explicit
symplectic methods are slightly adjusted as adaptive ones. The
adaptive methods have only two additional steps and a negligible
increase in computational cost as compared with the nonadaptive
ones. Their implementation is simple. Several dynamical
simulations of particles and photons near black holes have
demonstrated that the adaptive methods typically improve the
efficiency of the nonadaptive methods. Because of the desirable
property, the new adaptive methods are applied to investigate the
chaotic dynamics of particles and photons outside the horizon in a
Schwarzschild-Melvin spacetime. The new methods are widely
applicable to all curved spacetimes corresponding to Hamiltonians
or time-transformed Hamiltonians with the expected splits. Also
application to the backwards ray-tracing method for studying the
motion of photons and shadows of black holes is possible.

\end{abstract}


\emph{Unified Astronomy Thesaurus concepts}: Black hole physics
(159); Computational methods (1965); Computational astronomy
(293); Celestial mechanics (211)




\section{Introduction}
\label{sec:intro}

In recent several years, the shadow image of  the supermassive
black hole in the core of the elliptical galaxy M87* has been
detected by the Event Horizon Telescope (EHT Collaboration et al.
2019). The result is extremely consistent with that predicted by
Einstein's Theory of General Relativity. The black hole image is
formed due to the significant gravitational lensing effect  of
nearby radiation under the extremely strong gravity interaction of
the black hole. It is the lensed image at infinity of the photon
sphere, which corresponds to the boundary between photon orbits
escaping to infinity and photon orbits falling into the black hole
(Virbhadra 2024).

In fact, the black hole shadows are located at  the region in the
observer sky corresponding to photon orbits that spiral toward
into the black hole. In this sense, computations of black hole
shadows are attributed to studying the motion of photons outside
the horizons of black holes and finding the correspondence between
the falling light rays and the points in the sky. For integrable
spacetimes (e.g. the Kerr metric), analytical methods can
accurately provide the shadow edges like the photon sphere
(Bardeen $\&$ Cunningham 1973). However, no analytical methods but
numerical ones can carry out this task in the general cases (in
particular for nonintegrable spacetimes). Through numerically
backward integrations of the geodesic equations of the light rays
from the observer, one traces the evolution of the light rays and
identifies which of the light rays from the observer fall into the
center. This is  a backward ray-tracing method (Johannsen 2013).
Because the final evolutions of numerous points corresponding to
photon orbits in the sky are numerically traced, the computations
are hopelessly expensive. To save labor and to possess high
accuracy, an explicit integrator is necessary to employ large time
steps when an orbit runs far away from the black hole, but small
ones when the orbit is close to the horizon of the black hole.
That is, adaptive time steps with the desired stretching and
shrinking are needed in the numerical integration scheme. Due to
this advantage, adaptive time steps were added to fifth-order
Runge-Kutta schemes (Pu et al. 2016; White 2022), Adams-Moulton
method (Pelle et al. 2022), eight-order embedded Runge-Kutta
method (Kawashima et al. 2023) and fourth-order
Runge-Kutta-Fehlberg algorithm (Younsi et al. 2023) in these
ray-tracing codes.

The integration schemes adopted in the existing ray-tracing
methods as traditional solvers should have no problems in the
description of photon motions and black hole shadows when the
integration times are short. However, they would provide
unreliable results for long-term simulations because they lack the
constancy of the integrals of motion and the preservation of
symplectic structure which conservative Hamiltonian systems
possess. It is possible for a photon to slowly fall into a black
hole or to slowly scatter to infinity in some cases, e.g. the
presence of weak chaotical behavior in a nonintegrable spacetime.
Thus, long-term integration of the photon orbit is required for
the identification of the final state of the photon orbit. In
fact, many curved spacetimes can become Hamiltonian problems. The
most appropriate solvers for Hamiltonian systems are symplectic
methods which respect the inherent canonical nature of the
original Hamiltonian dynamics (Wisdom 1982; Ruth 1983; Feng 1986).
They show no secular errors in the integrals of motion such as the
conserved energy. They make the energy oscillatory but not be
exact conservation. An integrator with both symplecticity and
exact energy conservation is rare. Although an exactly
energy-conserving implicit integration scheme with the
preservation of the underlying Hamiltonian was given by Bacchini
et al. (2019a, b), it is not symplectic.

The Hamiltonian systems for curved spacetimes are not separable to
the variables and cannot be split into two terms that are
explicitly solvable. In this case, explicit symplectic integrators
seem to become useless. Instead, implicit symplectic integrators
can be applied for curved spacetimes (Brown 2006; Kop\'{a}\v{c}ek
et al. 2010; Seyrich $\&$ Lukes-Gerakopoulos 2012; Tsang et al.
2015). Semiexplicit symplectic methods for
inseparable Hamiltonian systems (Jayawardana $\&$ Ohsawa 2023;
Ohsawa 2023) can also be. They were developed along the extended
phase space method of  Pihajoki (2015), and are symplectic not
only in the extended phase space but also in the original phase
space. The explicit methods are less computationally demanding
than the implicit ones. Recently, our group showed that the
Hamiltonian systems of some curved spacetimes such as the
Schwarzschild metric can be separated into three or more
explicitly solvable parts (Wang et al. 2021a, b, c; Zhou et al.
2022, 2023). In this way, explicit symplectic methods are easily
implemented. However, some other curved spacetimes like the Kerr
metric have no such desirable splits. Wu et al. (2021, 2022) found
that the construction of explicit symplectic methods for these
curved spacetimes is still possible according to the idea of time
transformation considered by Mikkola (1997). Such
time-transformed explicit symplectic integrators were applied to
study the chaotic dynamics of particles near a magnetic deformed
Schwarzschild black hole (Huang et al. 2022). Constant step sizes
are adopted for the newly transformed time in the methods, and do
not break the symplectic property. Although the established
time-transformed explicit symplectic integrators allow for the
application of adaptive time steps to the original time, they have
some difficulty in the implementation due to a great deal of
computational cost. The time transformation functions for
constructing efficient explicit symplectic algorithms are close to
1 or constants in general when the distances are large enough.
This means that the original time and the new time are almost the
same, and the original time steps remain approximately constant.

Of course, the time-transformed explicit symplectic integrators
allow for the use of adaptive time steps in the work of  Mikkola
(1997). Such time transformation method is applied to improve the
efficiency of the symplectic leapfrog algorithm of Wisdom $\&$
Holman (1991) for highly eccentric orbits and close encounters
between objects in various few-body problems in the Solar System.
The logarithmic-Hamiltonian leapfrog algorithm is also an
efficient adaptive time step symplectic method for $N$-body
problems (Mikkola $\&$ Tanikawa 1999, 2013). A class of adaptive
time step, reversible, explicit symplectic integration algorithms
were formulated for the long-term integration of nearly Keplerian
orbits in separable Hamiltonian systems (Preto $\&$ Tremaine
1999). Emel'yanenko (2007) developed an explicit symplectic method
that allows individual adaptive time-steps in long-term
integrations of the planetary N-body problem. Following the
derivation of Emel'yanenko (2007), Preto $\&$ Saha (2009) provided
a symplectic scheme with adaptive time steps for  a post-Newtonian
Hamiltonian from the Kerr metric. This method is an explicitly and
implicitly combined algorithm. In the method, a auxiliary variable
$\Phi$ is used as a conjugate momentum with respect to the
original time being an additional coordinate. $\Phi$ is kept
constant during the calculation of the other coordinates and
momenta in every integration step, whereas it is advanced after
the step integration ends. In fact, $\Phi$ acts as a rescaled
factor for adjusting the time variables.

Combining the expected time transformation functions suggested by
Wu et al. (2021, 2022) and the  auxiliary variable $\Phi$
introduced by Preto $\&$ Saha (2009), we design an adaptive time
step explicit symplectic integrator for curved spacetimes. This is
the main attempt of the present paper. For the sake of this aim,
we describe the new adaptive time step explicit symplectic
integrator in Section 2. Then, the motions of particles and
photons around several black holes are used to show efficient
implementation of the new algorithm and to test the performance of
the new algorithm in Section 3. The applicability of the new
method to the chaotic dynamics of particles and photons  near a
Schwarzschild-Melvin black hole (Ernst 1976) is also considered.
Finally, the main results are concluded in Section 4. Possible
choices of time transformation functions for several curved
spacetimes are given in Appendix A.

\section{Adaptive time step explicit symplectic integrators in
curved spacetimes}

Consider that an axial-symmetric rotating black hole geometry in
the Boyer-Lindquist coordinates $(t,r,\theta,\phi)$ corresponds to
the covariant metric
\begin{eqnarray}
dS^{2} &=& g_{\alpha\beta}dx^{\alpha}dx^{\beta} \nonumber \\
&=& g_{tt}dt^2+2g_{t\phi}dtd\phi+g_{rr}dr^2 \nonumber \\
&& +g_{\theta\theta}d\theta^2+g_{\phi\phi}d\phi^2.
\end{eqnarray}
The metric components $g_{tt}$, $\cdots$, $g_{\phi\phi}$ are
functions of the two variables $r$ and $\theta$. This metric has
contravariant nonzero components
\begin{eqnarray}
g^{tt} &=& \frac{g_{\phi\phi}}{g_{tt}g_{\phi\phi}-g^{2}_{t\phi}},
~~~~
g^{t\phi} = \frac{g_{t\phi}}{g^{2}_{t\phi}-g_{tt}g_{\phi\phi}}, \nonumber \\
g^{\phi\phi} &=& \frac{g_{tt}}{g_{tt}g_{\phi\phi}-g^{2}_{t\phi}},
~~~~ g^{rr} = \frac{1}{g_{rr}}, ~~~~ g^{\theta\theta} =
\frac{1}{g_{\theta\theta}}. \nonumber
\end{eqnarray}

The geodesic motion of a massive test particle or massless photon
in this spacetime is described by a Hamiltonian system with two
degrees of freedom
\begin{eqnarray}
H = \frac{1}{2}g^{\alpha\beta}p_{\alpha}p_{\beta} = f(r,\theta)
+\frac{1}{2}g^{rr}p^{2}_{r}
+\frac{1}{2}g^{\theta\theta}p^{2}_{\theta},
\end{eqnarray}
where $f(r,\theta)$ is a function of $r$ and $\theta$ in the form
\begin{eqnarray}
f(r,\theta) =\frac{1}{2}(g^{tt}E^2 +g^{\phi\phi}L^2) -g^{t\phi}EL.
\end{eqnarray}
$p_{r}$ and $p_{\theta}$ are the covariant momenta with respect to
$r$ and $\theta$. The metric components do not explicitly depend
on $t$ and $\phi$, therefore, the specific energy $E$ and angular
momentum $L$ of the particle or photon are two constants of
motion, which correspond to the covariant momenta $p_t=-E$ and
$p_{\phi}=L$. Due to the particle's or photon's rest mass, the
Hamiltonian is also a constant of motion:
\begin{equation}
H=-p_0,
\end{equation}
where $p_0=1/2$ for the time-like geodesics but $p_0=0$ for the
null geodesics. Here, the speed of light $c$ and the gravitational
constant $G$ are taken as geometrized units, $c=G=1$.

\subsection{Comments on the existing time-transformed explicit
symplectic integrators without adaptive time steps}

Unlike the Hamiltonian systems of nonrotating black holes such as
the Schwarzschild black hole (Wang et al. 2021a), the Hamiltonian
system (2) for the rotating black holes like the Kerr black hole
is not directly split into several explicit integrable pieces in
general. This seems to show that explicit symplectic methods
become useless in the system (2). However, this problem may be
solved with the aid of the idea of time transformations for
constructing efficient symplectic algorithms proposed by Mikkola
(1997). Wu et al. (2021) found an appropriate time transformation
function to the Hamiltonian of Kerr geometry so that the obtained
time-transformed Hamiltonian has three or more splitting parts
which are explicitly solvable. Through such time transformations
and symplectic splitting methods, symmetric compositions can yield
explicit symplectic integrators. A possible path on the
construction of time-transformed explicit symplectic schemes for
the Hamiltonian (2) is introduced simply as follows.

Set $\tau$ as a physical time variable. It denotes the proper time
for the time-like geodesic motion, and is an affine time parameter
for the null  geodesic motion. Suppose that $w$ is a new time
variable different from the coordinate time $t$. The old time
$\tau$ and the new time $w$ satisfy the relation
\begin{equation}
d\tau=g(r,\theta)dw,
\end{equation}
where $g(r,\theta)$ is a time transformation function depending on
$r$ and $\theta$. This function should be chosen appropriately so
that the time-transformed Hamiltonian
\begin{eqnarray}
K &=& g(r,\theta)(H+p_0) \nonumber \\
&=&
g(r,\theta)(f(r,\theta)+p_0)+\frac{1}{2}g(r,\theta)g^{rr}p^{2}_{r} \nonumber \\
&& +\frac{1}{2}g(r,\theta)g^{\theta\theta}p^{2}_{\theta}
\end{eqnarray}
consists of several terms having analytical solutions as explicit
functions of the new time $w$. In fact, the second, third terms of
Equation (6) should be required to have the explicitly integrable
splits. $p_0$ in Equation (6) is the same as that in Equation (4)
and represents a momentum corresponding to a coordinate
$q_0=\tau$.\footnote{This means the old time $\tau$ treated as an
additional coordinate.} The Hamiltonian $H+p_0$ is based on the
use of the extended phase space. Because $H+p_0=0$, $K$ is always
identical to zero for any new time $w$, i.e. $K=0$. Via these
operations, the time-transformed Hamiltonian (6) has the splitting
expression
\begin{eqnarray}
K =\sum^{l}_{i=1}K_i.
\end{eqnarray}

Now, let symplectic operators $\psi_i$ be analytical solvers of
the sub-Hamiltonians $K_i$. Composing these symplectic operators
results in a first-order approximation to the exact solution of
the Hamiltonian (6):
\begin{eqnarray}
\chi(h) =\psi^h_l\times\cdots\times\psi^h_1,
\end{eqnarray}
where $h$ is a time step of the new time $w$. The operator $\chi$
corresponds to its adjoint
\begin{eqnarray}
\chi^*(h) =\psi^h_1\times\cdots\times\psi^h_l.
\end{eqnarray}
In terms of the two operators, a second-order explicit symplectic
method
\begin{eqnarray}
S_2(h)=\chi(\frac{h}{2})\times\chi^*(\frac{h}{2})
\end{eqnarray}
is symmetrically designed for the Hamiltonian (6) or (7). The
second-order method can easily compose higher-order explicit
symplectic integration algorithms (Yoshida 1990; Blanes $\&$ Moan
2002; Blanes et al. 2008, 2010;  Blanes et al. 2024). These
algorithms use fixed time steps in the new time, but may adopt
variable ones in the old time. This treatment is helpful to
preserve the symplectic structure of the Hamiltonian (6) in the
new time.

Clearly, how to find a suitable time transformation function and
how to split the time-transformed Hamiltonian are two keys for the
application of explicit symplectic integrators to the Hamiltonian
(2). In particular, the time transformation function is useful to
eliminate some factors in the second, third terms of the
inseparable Hamiltonian (2) so that the left second, third terms
satisfy the desired splitting needs. Some examples are listed
here. For the Kerr black hole with the spin angular momentum $a$,
Wu et al. (2021) gave the  time transformation function by
\begin{equation}
g(r,\theta)=\frac{\Sigma}{r^2}=\frac{1}{r^2}(r^2+a^2\cos^2\theta),
\end{equation}
and split the time-transformed Hamiltonian into five explicitly
integrable pieces. For other curved spacetimes such as the
rotating black ring, regular black holes, Gauss-Bonnet black hole,
majumdar-Papapetrou dihole black holes, relativistic core-shell
models and Kerr-Newman solution with disformal parameter, the time
transformation functions and splitting Hamiltonian methods were
also given by Wu et al. (2022). A notable point is that all the
time transformation functions chosen in the two works of Wu at al.
(2021, 2022) are nearly equal to 1 for sufficiently large radial
distances $r$, i.e. $g(r,\theta)\rightarrow 1$ as
$r\rightarrow\infty$. In other words, the old time step
$\Delta\tau$ is almost consistent with the new time step $\Delta
w=h$. As a result, both the old time $\tau$ and the new time $w$
have no typical differences. No adaptive time steps are employed
for the old time when a fixed time step is used for the new time.
This result was shown numerically in Figure 5b of Wu et al.
(2021).

In  principle, it should be admissible to replace the time
transformation function (11) with other functions, as was claimed
by Wu et al. (2021). For instance, the time transformation
function  is taken as
\begin{equation}
g_1=rg(r,\theta)=r+\frac{a^2}{r}\cos^2\theta,
\end{equation}
or
\begin{equation}
g_2=r^2g(r,\theta)=r^2+a^2\cos^2\theta.
\end{equation}
Seen from a theoretical point of view, the choice of the time
transformation function $g_1$ or $g_2$ can allow for the
application of explicit symplectic integrators in the Kerr
spacetime without doubt. In this case, the established explicit
symplectic methods like $S_2$ use adaptive time steps for the old
time. This point is illustrated here. When the separation $r$ is
large enough, $g_1\sim r$ and $g_2\sim r^2$; namely, $d\tau\approx
rdw$ for the use of $g_1$ and $d\tau\approx r^2dw$ for the use of
$g_2$. Thus, the original time-step selection is mainly controlled
by the radial distance $r$. For the use of constant new time step
$h$, the old time step $\Delta\tau$ is typically variable because
it decreases as a particle goes towards the black hole, whereas
increases when the particle runs away from the black hole. For a
small radial separation $r$, shortening of step sizes is useful to
improve the treatment of highly eccentric orbits or close
encounters. For a large radial separation $r$, enlarging of step
sizes is good to save computational cost. That is to say, the old
time steps $\Delta\tau$ have the desired stretching and shrinking.
Nevertheless, this does not lead to breaking the symplectic
property of the Hamiltonian (6) because the new time step $h$
remains fixed. In a word, improving accuracy and saving labors are
the benefits of adaptive time-step controls. Unfortunately, the
implementation of such  adaptive time step explicit symplectic
integrators for the choice of $g_1$ or $g_2$ is rather cumbersome
from practical computations because the smaller original time-step
selection requires choosing a much shorter new time step $h$, and
would lead to a great deal of computational cost as well as
numerous round-off errors.

Perhaps someone wants to know whether the performance of $S_2$
becomes better when the time transformation function $g_1$ gives
place to another form
\begin{equation}
g^\star=\frac{r}{j}g,
\end{equation}
where $j$ is a free parameter and $j\geq
r_{max}$\footnote{$r_{max}$ is the maximum value of $r$ in a
bounded orbit.} is possibly admitted. Although an appropriately
large new time step $h$ (e.g. $h=1$) can make the old time step
$\Delta\tau$ be extremely small in this case, the computations are
still more expensive and fail to yield putouts. A possible reason
for the algorithm $S_2$  not retaining the great speed advantage
of conventional symplectic integrators is that the factor $r/j$ in
Equation (14) is not only a rescaled time variable but also
affects the differential equations for the motion. If the role of
$r/j$ is only a rescaled time factor for adjusting the time-step
in integrations  and does not change the Hamiltonian canonical
equations of motion, then the adaptive time step explicit
symplectic integrator $S_2$ for the use of $g^\star$ would work
well. In what follows, we discuss the possibility of the function
$r/j$ that is only viewed as a rescaled time factor.

\subsection{Efficient implementation of new adaptive
time step explicit symplectic integrator}

As is mentioned above, the implementation of the highly efficient
explicit symplectic integrators such as $S_2$ requires that the
time transformation function $g$ of Equation (5) should satisfy
two points. One point is that the time transformation function can
successfully eliminate some factors impeding the second, third
terms of the Hamiltonian (2) so that the Hamiltonian (6) is split
into several explicitly integrable parts. The other point is that
$g\approx 1$ as the radial separation $r$ is sufficiently large.
On the other hand, Emel'yanenko (2007) derived an elegant
Hamiltonian of adaptive stepsize. Following the Hamiltonian
derivation, Preto $\&$ Saha (2009) developed a symplectic
integrator with adaptive time steps for fast and accurate
numerical calculation of a post-Newtonian Hamiltonian describing
the motion of a test particle in a Kerr metric. In the algorithmic
construction, they introduced a conjugate momentum $\Phi$ with
respect to the additional coordinate $\tau$ as a rescaled time
variable adjusting the time steps in integrations.\footnote{The
conjugate momentum of the coordinate $\tau$ is $p_0$ in Equation
(6). However, it is here that the conjugate momentum is $\Phi$ and
$p_0$ is only the constant 1/2 or 0 in Equation (4).} In addition,
the Keplerian part as the main part of the Hamiltonian is solved
analytically using the algorithmic regularization scheme of Preto
$\&$ Tremaine (1999), and the perturbation part consisting of
post-Newtonian and external potential terms is calculated by the
implicit midpoint method (i.e. a second order symplectic
integrator) of Feng (1986). Composing the explicitly analytical
solution and the implicitly numerical solution produces a
generalized leapfrog. This is the idea of Preto $\&$ Saha (2009)
about the construction of adaptive time step symplectic
integrator. Now, combining the expected time transformation
function $g$ and the  rescaled time variable $\Phi$, we design an
adaptive time step explicit symplectic integrator for the
Hamiltonian (2).

Transforming from the old time $\tau$ to a new independent time
variable $s$, i.e. $\tau\rightarrow s$ (the relation between the
two times is subsequently derived), we use $g$ and $\Phi$ to
readjust the Hamiltonian (2) or to slightly modify the Hamiltonian
(6) as
\begin{eqnarray}
F &=& \frac{K}{\Phi}+g\ln\left(\frac{\Phi}{\varphi(\tau)}\right)
\nonumber \\
&=& \frac{g}{\Phi}(H+p_0)+g\ln\left(\frac{\Phi}{\varphi(\tau)}\right) \nonumber \\
&=&
\frac{1}{\Phi}\sum^{l}_{i=1}K_i+g\ln\left(\frac{\Phi}{\varphi(\tau)}\right),
\end{eqnarray}
where $\varphi(\tau)=\varphi(r(\tau),\theta(\tau))$ is a given
function of the original time $\tau$.\footnote{$\varphi$ is a
function of $r$ and $\theta$, which are functions of $\tau$. Thus,
$\varphi$ is also a function that directly depends on $\tau$,
labeled as $\varphi=\varphi(r(\tau),\theta(\tau))=\varphi(\tau)$.}
If $g=1$, line 2 of Equation (15) is Equation (21) of Preto $\&$
Saha (2009).

When a set of solutions $r$, $\theta$, $p_r$ and $p_\theta$ are
obtained from the first term\footnote{These solutions also seem to
be connected with the second term because $g$ is a function of $r$
and $\theta$. Here $r$, $\theta$ and $\tau$ are regarded as three
independent coordinates, as they are in the Hamiltonian $H$.
However, $r$ and $\theta$ are directly related to $\tau$ in the
function $\varphi$. In addition, $\ln(\Phi/\varphi)=0$ will be
shown in a later discussion.} of line 3 in Equation (15), $\Phi$
remains constant. In this case, they are still given by the
algorithm $S_2(h)$ in Equation (10), but $h$ should be adjusted as
$h/\Phi$. Namely, $S_2(h/\Phi)$ provides the solutions $r$,
$\theta$, $p_r$ and $p_\theta$ at some time $s$. Note that $\Phi$
is viewed as a constant only in computing the solutions $r$,
$\theta$, $p_r$ and $p_\theta$, but varies with time $s$ after one
step integration according to one of Hamilton's equations for $F$:
\begin{eqnarray}
\frac{d\Phi}{ds} = -\frac{\partial F}{\partial \tau} =
g\frac{d}{d\tau}\ln\varphi.
\end{eqnarray}
In fact, it can also be derived from the sub-Hamiltonian
$F_1=-g\ln\varphi$. The evolution equation of $\tau$ is
\begin{eqnarray}
\frac{d\tau}{ds} = \frac{\partial F}{\partial \Phi} =
-\frac{g}{\Phi^2}(H+p_0)+\frac{g}{\Phi}=\frac{g}{\Phi}.
\end{eqnarray}
Here, $H+p_0=0$ is used. Equation (17) is also given by the
sub-Hamiltonian $F_2=g\ln\Phi$. It can give an explicitly
analytical solution to the time coordinate:
\begin{eqnarray}
\tau(s)=\tau_0 +s\frac{g}{\Phi}.
\end{eqnarray}
Combing Equations (16) and (17), we have the relation
\begin{eqnarray}
\frac{d}{d\tau}\ln\Phi = \frac{d}{d\tau}\ln\varphi,
\end{eqnarray}
equivalently,
\begin{eqnarray}
\frac{d}{d\tau}\ln\frac{\Phi}{\varphi} =0, ~or~
\frac{\Phi}{g}\frac{d}{ds}\ln\frac{\Phi}{\varphi} =0.
\end{eqnarray}
Hence, $\Phi/\varphi$ is constant, i.e. $\Phi/\varphi=C$. If $C=1$
at the starting time, then we have
\begin{eqnarray}
\Phi/\varphi=1
\end{eqnarray}
for any time $\tau$ or $s$. For simplicity, $\varphi$ is taken as
\begin{eqnarray}
\varphi=\frac{j}{r},
\end{eqnarray}
where  $j$ is still the free parameter in Equation
(14).\footnote{The simplest choice $\varphi\propto 1/r$ was
considered in the work of Preto $\&$ Saha (2009). We guess that
$\varphi=1/r$, i.e. Equation (22) with $j=1$, was used. Here, $j$
is not limited to 1 but has a widely free choice. In later
numerical simulations, we shall show what advantage exists for the
use of $j$ in the improvement of integration accuracy and what an
appropriate choice of $j$ is for an individual orbit in every
problem.} Substituting Equation (22) into Equation (16) yields
\begin{eqnarray}
\frac{d\Phi}{ds}= -\frac{g}{r}\frac{dr}{d\tau},
\end{eqnarray}
where $dr/d\tau=\partial H/\partial p_r=g^{rr}p_r$ is determined
by the Hamiltonian (2).  $\Phi$ is analytically solved by
\begin{eqnarray}
\Phi(s)=\Phi_0-s\frac{g}{r}g^{rr}p_r.
\end{eqnarray}
This shows that the parameter $j$ is independent of the
integration of $\Phi$ at time $s$, but is only dependent on the
initial value $\Phi_0$.

The above analyses demonstrate that the Hamiltonian $F$ is zero
for any time $s$ and consists of $l+2$ explicitly integrable
terms:
\begin{eqnarray}
F = \frac{1}{\Phi}\sum^{l}_{i=1} K_i+F_1+F_2.
\end{eqnarray}
It is obviously suitable for the application of an adaptive time
step explicit symplectic integrator of second order:
\begin{eqnarray}
AS_2(h) &=& \mathcal{F}_1(\frac{h}{2})\times
\mathcal{F}_2(\frac{h}{2})\times S_2(\frac{h}{\Phi}) \nonumber
\\ && \times\mathcal{F}_2(\frac{h}{2})\times\mathcal{F}_1(\frac{h}{2}),
\end{eqnarray}
where $\mathcal{F}_1$ and $\mathcal{F}_2$ respectively correspond
to the solvers of the sub-Hamiltonians $F_1$ and $F_2$, and
$h=\Delta s$ is still a fixed step size in the new time variable.
In what follows, several points on the adaptive explicit
symplectic scheme are illustrated.

Point 1: The basic clue for the application of the algorithm
$AS_2$ to the Hamiltonian (2) is concluded as follows.

1. Find the desired time transformation function $g$.

2. Split the Hamiltonian $K$ of Equation (6) in the expected form.

3. Write the Hamiltonian $F$ of adaptive stepsize in Equation
(15).

4. Apply a second order explicit symplectic method to solve the
Hamiltonian $F$.

Point 2: The algorithm $AS_2$ for $F$ is implemented as follows.

\emph{Step 1}: Advance $\Phi$ by $-ghg^{rr}p_r/(2r)$ in Equation
(24).

\emph{Step 2}: Advance $\tau$ by $gh/(2\Phi)$ in Equation (18).

\emph{Step 3}: Evolve $r$, $\theta$, $p_r$ and $p_\theta$ under
$K$ of Equation (6), using $S_2$ with the time interval $h/\Phi$
of Equation (10).

\emph{Step 4}: Reiterate step 2.

\emph{Step 5}: Reiterate step 1.

Point 3: Equations (17), (21) and (22) show that the relation
between the two times $\tau$ and $s$ is
\begin{eqnarray}
d\tau=g^{*}ds =\frac{g}{\Phi}ds=\frac{r}{j}gds,
\end{eqnarray}
where $g^{*}$ is another time transformation function. In
practice, $g^{*}$ is the time transformation function $g^{\star}$
of Equation (14). Although $g^{*}$ and $g^{\star}$ are the same
time transformation function, the factor $r/j$ has different
contributions to the computations of $r$, $\theta$, $p_r$ and
$p_\theta$ in the algorithms $S_2$ and $AS_2$. This factor similar
to $g$ typically affects the evolution equations of $r$, $\theta$,
$p_r$ and $p_\theta$ in the method $S_2$ with $g^{\star}$, and
then the computational efficiency is relatively poor. However,
$1/\Phi=r/j$ in Equation (17) is kept constant and only acts as a
scaling time factor adjusting the time steps in the computations
of $r$, $\theta$, $p_r$ and $p_\theta$ by $S_2(h/\Phi)$ of the
algorithm $AS_2$. Here, $\Phi$ remaining constant does not always
remain invariant for any time, but in fact it varies with time. It
does mean that only when all sub-Hamiltonians $K_i$ in Equation
(15) are solved, $\Phi$ is regarded as a constant (see Step 3 of
Point 2). After a step integration ends, $\Phi$ is advanced (see
Steps 1 and 5 of Point 2). In fact, many numerical experiments in
the previous works of Wang et al. (2021a, b, c) and Wu et al.
(2021) have confirmed that the integrator $S_2(h)$ is easy to
efficiently implement for the use of the desired time
transformation function $g$.\footnote{No time transformations are
needed in the works of Wang et al. (2021a, b, c). In other words,
$g=1$ is adopt.} In this case, $S_2(h/\Phi)$ should also
efficiently work as $\Phi$ remains constant. This result
reasonably supports that $AS_2(h)$ should be computationally more
efficient. The \emph{efficient implementation} of $AS_2(h)$ is due
to the \emph{constant of $\Phi$} in the course of solving all the
sub-Hamiltonians $K_i$. The function $g$ ($\approx 1$ for
sufficiently large $r$) in Equation (15) plays a key role in the
obtainment of explicitly integrable terms $K_i$.

Point 4: The use of different time steps in different parts of an
orbit leads to destroying the symplectic property of an integrator
that is symplectic for the use of a fixed step size. The fixed
time step $h$ is considered in the new time $s$, therefore, the
integrator $AS_2$ remains symplectic. Different time steps appear
in different parts of the orbit for the original time $\tau$.
Smaller old  time steps are used for the orbit in the vicinity of
the horizon of a black hole, while larger ones are for the orbit
far away from the black hole. This fact ensures higher accuracy of
the solution at the pericenter or a smaller separation $r$ and
higher efficiency at the apocenter or a larger separation $r$.
Adaptive time stepping can satisfy the needs. The desired
\emph{stretching and shrinking} of the old time steps $\Delta\tau$
are carried out via the \emph{advancing of $\Phi$} after every
integration step in the method $AS_2(h)$.

Point 5: The adaptive method $AS_2$ has only additional Steps 1
and 5 in Point 2 as compared with the nonadaptive method $S_2$.
Therefore, the former method needs a negligibly small amount of
cost of additional computation. The desired stretching and
shrinking of the step size resulting in the improvement of
efficiency of computations for $AS_2$ is based on both methods
achieving the same accuracy of computations. In this case, smaller
step sizes are needed.

Point 6: The adaptive time step explicit symplectic integrator
$AS_2$ is not limited to the application of the rotating spacetime
(1). In fact, it is suitable for any curved spacetimes, whose
corresponding Hamiltonians or time-transformation Hamiltonians
exist the desirable splits. A class of spacetimes with direct
symplectic analytical integrable decompositions and two sets of
spacetimes corresponding to appropriate time-transformation
Hamiltonians with the expected splits were introduced by Wu et al.
(2022). They allow for the use of $AS_2$. Obviously, the efficient
implementation of $S_2$ must lead to that of $AS_2$.

In a word, the adaptive time step explicit symplectic method is
simple to implement, low at the added cost of computation, and
widely applicable to many curved spacetimes.

\section{Tests and applications of the adaptive method}

The adaptive explicit symplectic method $AS_2$ is applied to
integrate three black hole models. In this way, we evaluate its
numerical performance by comparing it with the nonadaptive
explicit symplectic method $S_2$. The dynamical properties of
massive test particles or massless photons around some of the
black holes are focused on. In numerical tests, the mass of each
black hole takes one geometric unit, $M=1$. Dimensionless
variables, constants and parameters are also adopted.

\subsection{Schwarzschild black hole}

A dynamical model of charged particles near the Schwarzschild
black hole immersed into an external asymptotically uniform
magnetic field was used to test the nonadaptive explicit
symplectic method $S_2$ by Wang et al. (2021a). It is still
applied to check the adaptive explicit symplectic method $AS_2$.
The black hole metric components are
\begin{eqnarray}
g_{tt}&=& -\left(1-\frac{2}{r}\right),
~~~~g_{rr}=\left(1-\frac{2}{r}\right)^{-1}, \nonumber \\
g_{\theta\theta}&=& r^2, ~~~~~~~~~~~~~~~~~~~~
g_{\phi\phi}=r^2\sin^2\theta.
\end{eqnarray}
The magnetic field is given by a four-vector electromagnetic
potential having only one nonzero covariant component (Wald 1974)
\begin{eqnarray}
A_\phi = \frac{\tilde{B}}{2}r^2\sin^2\theta,
\end{eqnarray}
where $\tilde{B}$ is the strength of magnetic
field. The charged particle motion with charge $q$ is governed by
the Hamiltonian
\begin{eqnarray}
H &=& \frac{1}{2}g^{\mu\nu}(p_{\mu}-qA_{\mu})(p_{\nu} -qA_{\nu})
\nonumber \\
&=& -\frac{1}{2}(1-\frac{2}{r})^{-1}
E^{2}+(L-\frac{\beta}{2}r^{2}\sin^{2}\theta)^{2}\nonumber \\
&& /(2r^2\sin^2\theta)+\frac{1}{2}(1-\frac{2}{r})p^{2}_{r}
+\frac{1}{2}\frac{p^{2}_{\theta}}{r^2},
\end{eqnarray}
where $\beta=q\tilde{B}$. Take $r$ as the
dimensionless distance and $r_{pl}$ as the  practical distance.
Similar notations are given to the other quantities. The
dimensionless quantities and the practical ones have their
correspondences as follows:  $r_{pl}=Mr$, $t_{pl}=Mt$,
$\tau_{pl}=M\tau$, $E_{pl}=mE$, $q_{pl}=mq$, $p_{r,pl}=mp_r$,
$H_{pl}=mH$,\footnote{ If $H
=g^{\mu\nu}(p_{\mu}-qA_{\mu})(p_{\nu} -qA_{\nu})/(2m)$, then
$H_{pl}=mH$. However, $H_{pl}=m^2H$ if $H
=g^{\mu\nu}(p_{\mu}-qA_{\mu})(p_{\nu} -qA_{\nu})/2$.}
$\tilde{B}_{pl}=\tilde{B}/M$, $L_{pl}=MmL$ and
$p_{\theta,pl}=Mmp_{\theta}$. Here, $m$ is the mass of the charged
particle. Wang et al. (2021a) pointed out that this Hamiltonian
\emph{does not need any time transformation} and \emph{can
directly be split into four explicitly analytical solvable parts}.
The related quantities in Equations (6) and (7) correspond to
\begin{eqnarray}
p_0 &=& \frac{1}{2}, ~~~~~~~~ g(r,\theta)=1, \\
K_1 &=& f(r,\theta)+p_0 \nonumber \\
&=& -\frac{1}{2}(1-\frac{2}{r})^{-1} E^{2} +(L-\frac{\beta}{2}r^{2}\sin^{2}\theta)^{2} \nonumber \\
&& /(2r^2\sin^2\theta)+\frac{1}{2},
\\
K_{2} &=& \frac{1}{2}p^{2}_{r},\\
K_{3} &=& -\frac{1}{r}p^{2}_{r},\\
K_{4} &=& \frac{p^{2}_{\theta}}{2r^2}.
\end{eqnarray}
Hence, the methods $S_2$ and $AS_2$ can work.

The particle's energy $E=0.995$ and the angular momentum $L=4$ are
chosen. The new time step $h=1$ and the free parameter $j=100$ in
Equation (22) are used. Two cases without magnetic field $\beta=0$
and with magnetic field $\beta=9.7\times 10^{-4}$ are considered
in the Schwarzschild spacetime. An orbit with initial conditions
$r=7$, $\theta=\pi/2$, $p_r=0$ and $p_\theta>0$ solved from the
equation $H+p_0=0$ is tested. Figure 1a plots the phase space
structures of $r$ and $p_r$ of this orbit in the two cases via
Poincar\'{e} section at the plane $\theta=\pi/2$ with
$p_\theta>0$. The two algorithms $S_2$ and $AS_2$ are almost the
same in the obtained structures. In the case of  $\beta=0$, the
Schwarzschild spacetime is integrable and does not allow for the
presence of chaos. A regular Kolmogorov-Arnold-Moser (KAM) torus
on the Poincar\'{e} section is what we expect. For
$\beta=9.7\times 10^{-4}$, the Hamiltonian system (30) is
nonintegrable and possibly chaotic. Many discrete points
distributed in a area on the Poincar\'{e} section indicate the
characteristic of chaos. The chaotic behavior was also shown by
P\'{a}nis et al. (2019) using the time series analysis method.
Clearly, the distance  with $r_{max}\approx150$ for
$\beta=9.7\times 10^{-4}$ is smaller than that with
$r_{max}\approx190$  for $\beta=0$. That is, the existence of
magnetic field leads to decreasing the motion region of charged
particles, as compared with the absence of magnetic field.

For the two values of magnetic parameter $\beta$ in Figure 1b, the
methods $AS_2$  and $S_2$ give no secular growth to the errors of
the Hamiltonian, $\Delta H=H+p_0$. Such good long-term stability
and error behavior share some of the benefits of the standard
symplectic integrators with constant time steps. The errors in the
order of $10^{-7}$ for $AS_2$ are two orders of magnitude smaller
than those in the order of $10^{-5}$ for $S_2$. This shows that
the use of variable steps is favorable for the accuracy of
computations, as compared with that of constant steps. In
addition, the regular dynamics for $\beta=0$ and the chaotic
dynamics for $\beta=9.7\times 10^{-4}$ have no significant
differences in the Hamiltonian errors for each of the two
algorithms.

Seen from the relation between the proper time $\tau$ and the new
time $s$ for the method $AS_2$ in Figure 1c, the proper time
$\tau$ is about $9.7\times 10^{3}$ for the regular case of
$\beta=0$, and about $8.1\times 10^{3}$ for the chaotic case of
$\beta=9.7\times 10^{-4}$ when $s=10^{4}$. Namely, $\tau$ and  $s$
are nearly consistent for the former case, and are somewhat
different for the latter case.

Although the approximate equivalence of $\tau$ and  $s$ is present
in the method $AS_2$  for $\beta=0$, $AS_2$ still exhibits better
accuracy in the integrals of motion than $S_2$. This is attributed
to the application of the desired stretching and shrinking of the
proper time steps $\Delta\tau$ to $AS_2$. $\Delta\tau$ grows
nearly linearly with the distance $r$ in Figure 1d. Different step
sizes appear in different values of $r$ during integration of the
orbit. The step size is smaller for a shorter distance $r$,
whereas larger for a longer distance $r$. This  treatment of the
step sizes is useful to significantly improve accuracy and
efficiency of computations.

The accuracy of the method $AS_2$ relies on  different choices of
$j$ in Equation (22). The accuracy of integrations is improved
typically as $j$ increases in Table 1. The difference between
$\tau$ and $s$ gets larger, too. To make the difference as small
as possible, we roughly take $j\sim (r_{min}+r_{max})/2$, where
$r_{min}$ is the smallest distance of the bounded orbit. Based on
better accuracy and extremely smaller difference between $\tau$
and  $s$, $j=100$ is perhaps a better choice in the present
situation.

We have demonstrated in detail an efficient implementation of the
adaptive time step explicit symplectic integration algorithm to
the Schwarzschild spacetime. The implementation is relatively easy
because this spacetime has its Hamiltonian system that can be
directly split into several explicitly integrable terms without
time transformation (i.e. $g=1$). There are many other black hole
spacetimes satisfying this property. In fact, the known
spacetimes, including Reissner-Nordstr\"{o}m black hole (Wang et
al. 2021b), Reissner-Nordstr\"{o}m-(anti)-de Sitter black hole
(Wang et al. 2021c), modified gravity Schwarzschild black hole
(Yang et al. 2022), rotating black ring (Wu et al. 2022),
Brane-world metric (Hu $\&$ Huang 2022),
Einstein-\AE ther black hole (Liu $\&$ Wu 2023),
Konoplya-Zhidenko black hole (He et al. 2023),
and Horndeski gravity hairy black hole (Cao et al. 2024), have
been given the desired splitting forms. Without doubt, our
adaptive time step explicit symplectic scheme should also be well
suited for studies of these spacetimes.

\subsection{Kerr black hole}

The Kerr spacetime describes a rotating black hole geometry, which
has covariant nonzero components (Kerr 1963):
\begin{eqnarray}
&& g_{tt}=-(1-\frac{2r}{\Sigma}), ~~
g_{t\phi}=-\frac{2ar\sin^{2}\theta}{\Sigma}=g_{\phi t}, \nonumber \\
&&  g_{rr}=\frac{\Sigma}{\Delta}, ~~~~~~~~~~~~
g_{\theta\theta}=\Sigma,  \nonumber \\
&& g_{\phi\phi}=(\rho^2+\frac{2ra^2}{\Sigma}
\sin^{2}\theta)\sin^{2}\theta,
\end{eqnarray}
which $\Delta=\rho^2-2r$ and $\rho^2=r^2+a^2$.
The dimensionless spin parameter $a$ corresponds
to its practical spin  $a_{pl}=aM$. For the Kerr geometry, the
Hamiltonian (2) cannot be directly split into more than two
explicitly integrable parts because $\Sigma$ acts as the
denominators of the second, third terms of Equation (2). However,
Wu et al. (2021) can still split the time-transformed Hamiltonian
(6) into five explicit analytical solvable terms using the time
transformation function $g$ of Equation (11) to eliminate the
factor $\Sigma$ of the denominators. This brings a good chance for
the application of the method $S_2$. Naturally, the method $AS_2$
is easily available, too.

\subsubsection{Time-like geodesics}

For the time-like geodesic motion of a massive test particle
around the Kerr black hole, we take $p_0=1/2$ in Equations (6) and
(15). Suppose that the black hole spin parameter is $a=0.8$, and
the particle has its energy $E=0.995$ and angular momentum
$L=4.6$. The new time step $\Delta s=h=1$ and $j=100$ are still
used. An orbit with its initial conditions of $r=9.8$, $p_r=0$ and
$\theta=\pi/2$ is chosen as a tested orbit. It is regular because
of the integrability of the Kerr spacetime.

The accuracy in the Hamiltonian $K$ of Equation (6) is two orders
of magnitude better for $AS_2$ than for $S_2$, as shown in Figure
2a. However, the proper time $\tau$ for  $S_2$ or $AS_2$ arrives
at 10,000 when  the time $w$ or $s$ reaches 10,000. That is,
$\tau\approx w$ for  $S_2$ and $\tau\approx s$ for  $AS_2$. Of
course, there are some differences between them in Figure2b.
$\tau$ increases linearly with time $w$ for $S_2$, whereas it does
nonlinearly with time $s$ for $AS_2$. $j=100$ for $AS_2$ is again
demonstrated to be an appropriate choice adjusting the relation
between $\tau$ and $s$ as the smallest difference. In spite of no
typical differences between the proper time $\tau$ and the
transformation time $w$ or $s$ in the two methods, the proper time
step $\Delta \tau$ of $S_2$ almost remains constant, while that of
$AS_2$ is variable in Figure 3c. The constant step size of $S_2$
is adopted due to the time transformation function $g\approx 1$ of
Equation (11) for sufficiently large distance $r$. The adaptive
time steps of $AS_2$ are employed because of $\Phi$ in the time
transformation function $g^{*}=g/\Phi\approx1/\Phi$ of Equation
(27) being varied according to Equation (24). The use of adaptive
time steps makes a crucial contribution to the accuracy of $AS_2$
superior to that of $S_2$. In addition to the dependence of the
proper time steps $\Delta \tau$ on the distances $r$ seen from
Figure 3c, the integrated orbit is bounded because its distances
$r$ have the maximum value and the minimum value.

\subsubsection{Null geodesics}

For the null geodesic dynamics of massless photons near the Kerr
black hole, we take $p_0=0$ in Equations (6) and (15). Unlike the
orbits of particles, all orbits of photons are unstable except
some particular regions. For example, the black hole regime
$0<r<r_{h^{-}}$ within the inner horizon $h^{-}=1-\sqrt{1-a^2}$
($|a|<1$) allows the existence of stable equatorial circular
photon orbits (Dolan $\&$ Shipley 2016). Photon orbits outer the
horizons of the Kerr black hole belong to one of the three
regions: falling to the center, scattering to infinity and
unstably circling in the center (Bardeen $\&$ Cunningham 1973).

Consider that a photon orbit scattering to infinity corresponds to
the parameters $a=0.3$, $E=0.99$, $L=3\sqrt{3}E/10$ and the
initial conditions $r=3$, $\theta=\pi/2$, $p_r=0$. Taking $h=0.1$
and $j=1000$, we integrate this escaping orbit until the distance
increases to $r=1000$. Although the two methods $S_2$ and $AS_2$
almost run the same original time $\tau\approx1,016$, they use
different new times: $s=1,016$ for  $S_2$ and  $s=8,103$  for
$AS_2$. Because of this point, the curves $\tau$-$K$ for two
algorithms are plotted in Figure 3a instead of the curves $w$
$(s)$-$K$ in Figure 2a. The errors in the Hamiltonian $K$ still
remain stable. They are about two orders of magnitude larger  for
$S_2$ than  for $AS_2$. In Figure 3b, the old step sizes are
basically constant for $S_2$ but grow linearly with the distances
$r$ for  $AS_2$. Different values of $j$ in the method $AS_2$
affect the accuracies of $K$ and the values of new time $s$. Seen
from Table 2, $j=1000$ is perhaps the best choice.

What about the performance of the method $AS_2$ when this
scattering orbit is replaced with an orbit falling to the black
hole? To answer this question, we choose the parameters $a=0.5$,
$E=0.996$, $L=-2.211$ and the initial conditions $r=100$,
$\theta=\pi/2$, $p_r=-1.016$. $h=0.001$ is used, and several
values are also given to $j$ in the integrator $AS_2$. The
integrations have to end when the distance $r$ decreases to 2 at
the nearby horizon of the black hole. Figure 4a shows that the
increase of $j$ does not necessarily lead to good results, and
$j=100$ should be the best choice. The symplectic integrators
yield drifts in the errors $K$ when the orbit approaches the
horizon. $AS_2$ with $j=100$ can perform better accuracy in the
vicinity of the horizon than $S_2$. This advantage is helpful to
apply $AS_2$ to a new ray-tracing method to obtain black hole
images by finding the related photon regions in the observer's
plane. Both the fixed time step of $S_2$ and the adaptive time
steps of $AS_2$ are displayed clearly in Figure 4b.

In short, the simulations of the motions of particles and photons
around the Kerr black hole show that the efficient implementation
of $AS_2$ has two keys. One key is the obtainment of the desired
time transformation functions $g$. $g$ should make the related
time-transformed Hamiltonians have explicitly integrable splits,
and should be approximately equal to 1 or some constants. Such
functions can be found in some other literature besides the works
of Wu et al. (2021), Sun et al. (2021a, b) and Wu et al. (2022) on
the Kerr type spacetimes. They exist in non-Schwarzschild black
holes in modified theories of gravity (Zhang et al. 2021), regular
charged black hole (Zhang et al. 2022),
Reissner-Nordstr\"{o}m-Melvin black hole (Hu $\&$
Huang 2023; Yang et al. 2023), Kerr-Newman-Melvin black hole
(Yang $\&$ Wu 2023) and renormalized group improved Schwarzschild
black hole (Lu $\&$ Wu 2024). In addition to these examples, the
desired time transformation functions $g$ in several black hole
spacetimes are listed in Appendix A. It is also easy to find such
similar time transformation functions $g$ when the method $AS_2$
is applied to simulate the chaotic dynamics of particle orbits
around Kerr black holes surrounded by external magnetic fields
(Takahashi $\&$ Koyama 2009; Stuchl\'{i}k $\&$ M. Kolo\v{s} 2016;
Tursunov et al. 2016; Kop\'{a}\v{c}ek $\&$ Karas 2018; Tursunov et
al. 2020; Kolo\v{s} et al. 2021). The other key is an appropriate
choice of $j$. In general, $j$ is roughly given in the range of
$(r_{min}+r_{max})/2\leq j\leq r_{max}$. In particular, it would
be good to choose $j$ as the observer's distance in ray-tracing
integrations.

\subsection{Schwarzschild-Melvin black hole}

Ernst (1976) gave a Schwarzschild-Melvin black hole describing the
Schwarzschild black hole immersed in the magnetic universe of
Melvin (1965). This black hole is an exact solution of Einstein's
equations. Besides the solution, the Reissner-Nordstr\"{o}m-Melvin
black hole and Kerr-Newman-Melvin black hole were presented. The
Schwarzschild-Melvin black hole has nonzero metric components
\begin{eqnarray}
&& g_{tt}=-\Lambda^2 \left(1-\frac{2}{r}\right), ~~
g_{rr}=\Lambda^2 \left(1-\frac{2}{r}\right)^{-1}, \nonumber \\
&& g_{\theta\theta}=\Lambda^2r^2,  ~~~~~~~~~~~~~~~~
g_{\phi\phi}=\Lambda^{-2}r^2\sin^{2}\theta,
\end{eqnarray}
where function $\Lambda$ is given by
\begin{eqnarray}
\Lambda=1+\frac{1}{4}B^2r^2\sin^{2}\theta.
\end{eqnarray}
The dimensionless magnetic field $B$ corresponds
to the practical one $B_{pl}=B/M$.

If the magnetic field vanishes, this spacetime is the
Schwarzschild black hole. When it is nonzero, it acts as a
gravitational effect, which causes this spacetime not to be
asymptotically flat. If $|B|\ll1$ and $2\ll r\ll 1/B$, then
$\Lambda\approx 1$ outside the horizon. This means the existence
of the approximately flat spacetime and the approximately uniform
magnetic field. In this case, this spacetime is \emph{nearly
integrable}. This fact was confirmed by finding chaos of neutral
particles around the Schwarzschild-Melvin black hole in the works
of Karas $\&$ Vokrouhlick\'{y} (1992) and Li $\&$ Wu (2019). Of
course, neutral particles can also be chaotic in the
Reissner-Nordstr\"{o}m-Melvin black hole spacetime (Yang et al.
2023) and Kerr-Newman-Melvin black hole spacetime (Yang $\&$ Wu
2023). On the other hand, self-similar fractal structures appear
in the shadows of the Schwarzschild-Melvin black hole (Junior et
al. 2021) and Kerr-Melvin black hole (Wang et al. 2021; Hou et al.
2022). The chaotic lensing, i.e. the chaotic motion of photons, is
responsible for these self-similar fractal structures in the black
hole shadows.

We select the time transformation function of Equation (5) as
\begin{eqnarray}
g=\Lambda^2.
\end{eqnarray}
The time-transformed Hamiltonian $K$ of Equation (6) resembles the
Hamiltonian $H$ of Schwarzschild black hole in Equation (30) with
$\beta=0$. That is, it consists of four explicitly integrable
terms. In fact, only Equation (32) in Equations (32)-(35) is
slightly modified as
\begin{eqnarray}
K_1 = -\frac{rE^2}{2(r-2)}
+\frac{L^{2}\Lambda^4}{2r^2\sin^2\theta} +p_0\Lambda^2.
\end{eqnarray}
Although $E$ and $L$ are usually called as the energy and angular
momentum of the photon in the Schwarzschild-Melvin spacetime, they
are unlike those in the Kerr metric. For the former spacetime, $E$
is not the particle's or photon's energy measured by an observer
at spatial infinity unless the observer lies on the axis of
symmetry (Junior et al. 2021). However, $E$ is always the energy
measured at spatial infinity for the latter spacetime. It is clear
that the methods $AS_2$ and $S_2$ are suitably applicable to the
time-transformed Hamiltonian $K$ of the Schwarzschild-Melvin black
hole spacetime.

\subsubsection{Dynamics of massive test particles}

For the motion of a massive neutral test particle around the black
hole, the parameters are given by $E=0.9905$, $L=3.6$ and
$B=0.001$. The step size $h=1$ and the parameter $j=100$ are also
given. In Figure 5a, $g\approx 1$ outside the horizon  can be
guaranteed obviously. The method $S_2$ seems to show the
regularity of Orbit 1 with the initial separation $r=10$. However,
the method $AS_2$ describes that the orbit allows the onset of
chaos confined to very narrow bands in this section. The result is
consistent with that of Figure 4b in the work of Li $\&$ Wu
(2019). The chaoticity of such a neutral particle is due to the
magnetic field in Equation (38) acting as a gravitational effect
in the spacetime geometry. Notice that the magnetic fields in
Equations (29) and (38) have different contributions. The magnetic
field in Equation (29) does not affect the spacetime background
(28) but the motion of charged particles in Equation (30). Namely,
the spacetime (28) is still integrable, but the Hamiltonian (30)
is not. Nevertheless, the magnetic field in Equation (38) directly
changes the spacetime background (37) and makes it nonintegrable.
Naturally, it affects the motion of particles. In this sense, the
occurrence of chaos is possible. Both methods almost give the same
dynamical information to any one of Orbits 2-6. In particular,
Orbit 5 with the initial separation $r=48.7$ has a saddle point
with one stable direction and another unstable direction.

Both algorithms showing different results on the dynamics of Orbit
1 is because $AS_2$ provides two orders of magnitude better
accuracy than $S_2$ in Figure 5b. When the integration adds up to
10,000 steps, the proper time $\tau$ is approximately equal to the
time $w$ for $S_2$, and about half the time $s$  for $AS_2$ in
Figure 5c. The nearly constant step size $\Delta\tau$ is used for
$S_2$, while the variable step sizes $\Delta\tau$ are for $AS_2$
in Figure 5d.

The method $AS_2$ is used to explore the transition from order to
chaos as the magnetic field strength $B$ is varied. When the
magnetic field strength is slightly smaller than that in Figure 5,
all orbits in Figure 5a become regular for $B=0.8\times 10^{-3}$
in Figure 6a. The shapes of Orbits 2-6 are unlike those in Figure
5a. A saddle point exists for Orbit 5 in Figure 5a, but does not
exist in Figure 6a. As the magnetic field strength is slightly
larger than that in Figure 5, chaos of Orbit 1 in Figure 5 dies
out for $B=1.2\times 10^{-3}$ in Figure 6b. Orbit 5 and orbit 7
with the initial distance $r=45$ become chaotic in the present
case. The chaotic region in Figure 6b is larger than in Figure 5a.

It can be seen from Figures 5 and 6 that an increase of the
magnetic field strength enhances the extent of chaos of massive
neutral test particles. Strengthening the extent of chaos is
considered from a statistical result of the global phase space
rather than  one single orbit.

\subsubsection{Dynamics of massless photons}

Now, we use the method $AS_2$ with the step size $h=1$ to simulate
the motion of a photon near the Schwarzschild-Melvin  black hole.
The parameters are $E=0.995$, $B=0.1$ and $j=100$. Consider that
an orbit has its initial radius  $r=15$. The angular momentum is
governed by $L=-E\sqrt{-g_{\phi\phi}/g_{tt}}$.\footnote{The choice
of $L$ is based on the impact parameter $\eta=L/E$ for the
effective potential vanishing at the equatorial plane. Here, the
potential is nonzero for the initial radius  $r=15$.} This orbit
is chaotically filled in a great region of Figure 7a. When the
parameters $E$ and $L$ are given, a slight decrease of the
magnetic field strength $B$ causes the transition from chaos to
order. The chaotic regions become gradually small and the number
of regular KAM tori increases for the magnetic field strength
varied from $B=0.095$ in Figure 7b to $B=0.091$ in Figure 7c and
$B=0.088$ in Figure 7d. When $B=0.086$ in Figure 7e and $B=0.084$
in Figure 7f, chaos is absent and all orbits are regular KAM tori.

In fact, Figure 7 shows the existence of stable bound regular
photon orbits and stable bound chaotic photon orbits, which
neither fall into the black hole nor escape to infinity. This
figure is similar to Figures 5 and 6 for the description of stable
bound regular particle orbits and stable bound chaotic particle
orbits existing in the Schwarzschild-Melvin spacetime. It is
unlike Figures 3 and 4, where no stable bound photon orbits but
falling photon orbits and escaping ones exist outside the horizon
in the Kerr spacetime. These results are easily shown via
effective potentials on the equatorial plane. The photon effective
potential similar to the particle one for the Schwarzschild-Melvin
spacetime has closed pockets corresponding to minimum values
(Junior et al. 2021). The presence of such closed pockets allows
that of stable bound photon orbits on (or outside) the equatorial
plane in the Schwarzschild-Melvin spacetime. In a similar way,
stable light rings could exist in the Kerr-Melvin spacetime (Wang
et al. 2021d). There are also stable light rings and chaotic
lensing of light near rotating boson stars and Kerr black holes
with scalar hair (Cunha et al. 2015, 2016). Stable regular or
chaotic photon orbits were found in stationary axisymmetric
electrovacuum (Dolan $\&$ Shipley 2016). However, the absence of
closed pockets in the photon effective potential does not allow
for the existence of stable bound photon orbits outside the
horizon of the Kerr spacetime.

Comparing between Figure 7 and Figures 5, 6, one clearly finds
that the stable bound photon orbits can exist for larger magnetic
fields in the Schwarzschild-Melvin spacetime, but the stable bound
particle orbits can for smaller magnetic fields. These results are
also seen from the effective potentials on the equatorial plane.
Wang et al. (2021d) showed that the radii of stable light rings
decrease with the magnetic field parameter $B$ increasing in the
Kerr-Melvin spacetime. For the Kerr spacetime with $B=0$, the
stable light ring radius tends to infinity. This point means that
no stable light rings exist outside the horizon in the Kerr
spacetime. The effective potentials for the Schwarzschild-Melvin
spacetime are also similar to those for the Kerr-Melvin spacetime.
The stable particle circular orbits can be allowed for smaller
magnetic fields, whereas the smaller radii of stable light rings
can for larger magnetic field parameters $B$. The smaller radii of
stable light rings mean the shapes of the effective potentials
going towards the black hole and an increase of the gravitational
effects. This result leads to enlarging the motion regions of
stable bound photon orbits and strengthening the extent of chaos
outside the equatorial plane for larger magnetic field strengths
in the Schwarzschild-Melvin spacetime.

Because a larger value of the magnetic field strength $B$ can
easily allow the presence of stable bound orbits outside the
equatorial plane in the Schwarzschild-Melvin spacetime,
$\Lambda\approx 1$ is no longer present. In fact, the range of
$\Lambda$ in Figure 7a is $1<\Lambda <1.6$, which is different
from $\Lambda\approx 1$ in Figures 5 and 6. It is clearly shown in
Figure 8a that the range of $g$ is $1.2<g<2.5$. In other words,
the old time steps $\Delta\tau$ for the method $S_2$ roughly use
variable step sizes in the range of $1.2<\Delta\tau<2.5$. The old
time steps for the method $AS_2$ are adjusted according to the
variation of $r$ from 0.1 to 1.1. The old time steps $\Delta\tau$
do not grow linearly with the separation $r$ for both methods. In
this case, the differences between the old times $\tau$ and the
time-transformed times $w$ or $s$ should be large in Figure 8b.
After $10^5$ integration steps, the old times are  $\tau=1.8\times
10^5$ for $S_2$, and $\tau=3\times 10^4$ for $AS_2$. The error in
the Hamiltonian $K$ is about two orders of magnitude smaller for
$AS_2$ than for $S_2$, as shown in Figure 8c. When the chaotic
orbit in Figure 7a is replaced with the regular orbit with the
initial distance $r=15$ in Figure 7b, the Hamiltonian errors
between the two cases have no explicit differences for each of the
two algorithms.

\section{Summary}

Some curved spacetimes, which are directly split into several
explicitly integrable parts, are easily fit for the application of
explicit symplectic integrators, as was introduced in the previous
papers (Wang et al. 2021a, b, c). Some other curved spacetimes,
which are not directly split into several explicitly integrable
parts but are through appropriate time transformations, can also
allow for the construction of explicit symplectic methods, as was
reported in the previous works (Wu et al. 2021, 2022). Adaptive
time steps should be admitted in such time-transformed explicit
symplectic methods from the theoretical viewpoint. However, they
are rather cumbersome to implement in general from practical
computations. The time transformation functions for the efficient
implementation of time-transformed explicit symplectic algorithms
should satisfy the time-transformed Hamiltonians having explicitly
integrable splits, and should approach 1 or some
constants\footnote{This requirement is considered in the generic
case. There are exceptional cases, e.g. the choice of time
transformation function for the motion of photons around the
Schwarzschild-Melvin  black hole.} for sufficiently large radial
distances.

Combing the desired time transformation functions and  the
step-size control technique proposed by Preto $\&$ Saha (2009), we
have developed an adaptive time step explicit symplectic
integrator for many curved spacetimes. Besides an appropriate
choice of the time transformation function, another key for the
efficient implementation of such an adaptive time step explicit
method is the introduction of a conjugate momentum $\Phi$ with
respect to the  old time as an additional coordinate. Here $\Phi$
acts as a rescaled time variable adjusting the time steps in
integrations. The constant of $\Phi$ in the course of solving the
other momenta and coordinates brings the efficient implementation
and good computational efficiency of the proposed algorithm. The
advancing of  $\Phi$ after every integration step leads to the
desired stretching and shrinking of the old time steps in the
method. A suitable choice of the parameter $j$ is also necessary.
The adaptive method has only two additional steps as compared with
the nonadaptive one. The efficient implementation of the latter
method naturally results in that of the former one. Therefore, the
adaptive method needs a negligibly small amount of cost of
additional computation, and its implementation is simple.

We have shown that the new adaptive time step explicit symplectic
integrator works well for numerical studies of several dynamical
problems. These problems include the motion of particles around
the Schwarzschild black hole immersed in an external magnetic
field, the dynamics of particles and photons near the Kerr black
hole, and the motion of particles and photons around the
Schwarzschild-Melvin black hole. The numerical tests have
demonstrated that the adaptive method typically improves the
efficiency of the  nonadaptive method. Even if the original time
and the newly transformed time have no explicit differences (see
e.g. Figures 1c, 2b and Tables 1, 2), the desired stretching and
shrinking of the stepsize still show good performance in the
suppression of long-term qualitative errors. Because of  such
desirable property,  the new adaptive method is applied to
investigate dependence of the transition from order to chaos on a
small change of the magnetic field strength. It is found that
stable bound photon orbits involving regular orbits and chaotic
ones, which neither spiral toward  into the black hole nor spiral
out to infinity, can exist outside the horizon in the
Schwarzschild-Melvin spacetime, but cannot in the Kerr spacetime.
In addition, the chaotic regions are enlarged with an increase of
the magnetic field strength for either the motion of particles or
the motion of photons around the Schwarzschild-Melvin black hole.

The adaptive time step explicit symplectic method
is suitable for integrations of highly eccentric
orbits and close encounters in the Solar System. In addition, it
can be applied to  many curved spacetimes ranging from the
problems considered in the present paper, the spacetimes motioned
in Appendix A and the literature (e.g. Wang et al. 2021a; Wu et
al. 2021, 2022; Yang $\&$ Wu 2023; Cao et al. 2024), and some
other spacetimes such as scalar fields (Virbhadra et al. 1998),
black-bounce- Reissner -Nordstr\"{o}m spacetime (Zhang $\&$ Xie
2022) and quantum gravity Schwarzschild black holes (Gao $\&$ Deng
2021; Huang $\&$ Deng 2024). It is useful to
simulate the motion of particles and photons outside the black
hole horizons in the phase space. In particular,
it is suitable for integrations of particles and
photons in the vicinity of the black hole horizons.
It is also applicable to backwards ray-tracing
methods, which study the light rays from points in an observer's
local sky how to correspond to the final evolution of photon
orbits in the phase space and obtain shadows of black holes. An
appropriate choice of the parameter $j$ of Equation (22) would be
the observer's distance in the ray-tracing integrations.

\appendix

\section{Suitable choices of time transformation
functions $g$ in several black hole spacetimes}

Four black hole spacetimes are given, and the related time
transformation functions are also provided.

\subsection{Einstein-Maxwell-Dilaton-Axion metric}

The Einstein-Maxwell-Dilaton-Axion metric describes a static,
axisymmetric rotating black hole, as a non-general relativity
black hole solution coupling metric parameters to a physical
field. Garc\'{i}a et al. (1995) gave the metric components:
\begin{eqnarray}
&&
g_{tt}=-\left(\frac{\hat{\Delta}-a^2\sin^2\theta}{\hat{\Sigma}}\right),
~~~~~~~~~~~~
g_{t\phi}=-\frac{a(\delta-W\hat{\Delta})\sin^{2}\theta}{\hat{\Sigma}}, \nonumber \\
&&  g_{rr}=\frac{\hat{\Sigma}}{\hat{\Delta}}, ~~~~~~~~~~~~
g_{\theta\theta}=\hat{\Sigma}, ~~~~~~~~~~~~
g_{\phi\phi}=\frac{\hat{A}}{\hat{\Sigma}} \sin^{2}\theta,
\end{eqnarray}
where the related notations are expressed as
\begin{eqnarray}
W &=&
1+[\beta_{ab}(2\cos\theta-\beta_{ab})+\beta^2_{a}]\csc^2\theta,
\\
\hat{\Sigma} &=&
\Sigma-(\beta^2+2br)+\beta_b(\beta_b-2a\cos\theta), \\
\hat{\Delta} &=& \Delta-(\beta^2+2br)+(1+2b)\beta^2_{b}, \\
\hat{A} &=& \delta^2-a^2\hat{\Delta}W^2\sin^{2}\theta, \\
\delta &=& r^2-2br+a^2.
\end{eqnarray}
$b$ and $\beta$ with  units of length are the coupling parameters
of the dilaton and axion fields. $\beta_a$, $\beta_{ab}$ and
$\beta_b$ are dimensionless parameters. This metric is also found
in the paper of Younsi et al. (2023).

Noting Equation (2), we obtain
\begin{eqnarray}
 g^{rr}=\frac{\hat{\Delta}}{\hat{\Sigma}}, ~~~~~~~~~~~~
g^{\theta\theta}=\frac{1}{\hat{\Sigma}}.
\end{eqnarray}
When the time transformation function of Equation (5) is taken as
\begin{eqnarray}
 g=\frac{\hat{\Sigma}}{r^2},
\end{eqnarray}
the Hamiltonian (6) is made of explicitly solvable five terms, and
$g\rightarrow 1$ for $r\rightarrow\infty$.

\subsection{Charged Horndeski black hole}

An asymptotically flat solution describing a charged Horndeski
black hole is shown in the paper of  Cisterna $\&$ Erices (2014)
as follows:
\begin{eqnarray}
&&
g_{tt}=-\left(1-\frac{2}{r}+\frac{4Q^2}{r^2}-\frac{4Q^4}{3r^4}\right),
~~~~~~~~~~~~ g_{\theta\theta}=r^2, \nonumber \\
&&  g_{rr}=-\frac{(8r^2-16Q^2)^2}{64r^4g_{tt}},
~~~~~~~~~~~~~~~~~~~~~~~~~~~ g_{\phi\phi}=r^2 \sin^{2}\theta,
\end{eqnarray}
where $Q$ is the electric charge parameter of black hole. Letting
the time transformation function  be
\begin{eqnarray}
 g=-\frac{(8r^2-16Q^2)^2}{64r^4g_{tt}},
\end{eqnarray}
we obtain $g\rightarrow 1$ for $r\rightarrow\infty$ and can split
the Hamiltonian (6) into several explicitly solvable pieces.

\subsection{Hairy Kerr black hole}

A stationary and axisymmetric hairy Kerr black hole is given by
Contreras et al. (2021) in the form
\begin{eqnarray}
&& g_{tt}=-\frac{\check{\Delta}-a^2\sin^{2}\theta}{\Sigma},
~~~~~~~~~~~~~~~~
g_{t\phi}=-a\left(1-\frac{\check{\Delta}-a^2\sin^{2}\theta}{\Sigma} \right)\sin^{2}\theta, \nonumber \\
&&  g_{rr}=\frac{\Sigma}{\check{\Delta}}, ~~~~~~~~~~~~
g_{\theta\theta}=\Sigma, ~~~~~~~~
g_{\phi\phi}=\sin^{2}\theta\left(\Sigma+a^2\sin^{2}\theta
\left(2-\frac{\check{\Delta}-a^2\sin^{2}\theta}{\Sigma} \right)
\right),
\end{eqnarray}
where
\begin{eqnarray}
\check{\Delta}=r^2+a^2-2r+\xi r^2 e^{-r/(1-l_0/2)}.
\end{eqnarray}
Here $\xi$ is a deviation parameter, and $l_0$ is a primary hair
parameter. If the time transformation function  is chosen as
\begin{eqnarray}
 g=\frac{\Sigma}{\check{\Delta}},
\end{eqnarray}
then $g\rightarrow 1$ for sufficiently large distance $r$ and the
Hamiltonian (6) has the desired split.

\subsection{Loop quantum gravity rotating black hole}

Islam et al. (2023) investigated Event Horizon Telescope
observations of the effects of a rotating black hole solution in
loop quantum gravity (Kumar et al. 2022). This solution is
described by
\begin{eqnarray}
&&
g_{tt}=-\left(1-\frac{2\bar{M}}{\bar{\rho}^2}\sqrt{r^2+l^2}\right),
~~~~~~~~~~~~
g_{t\phi}=-2a\bar{M}\sqrt{r^2+l^2}\sin^{2}\theta, \nonumber \\
&&  g_{rr}=\frac{\bar{\rho}^2}{\bar{\Delta}}, ~~~~~~~~~~~~
g_{\theta\theta}=\bar{\rho}^2, ~~~~~~~~~~~~~~~~~~~
g_{\phi\phi}=\frac{\bar{A}}{\bar{\rho}^2} \sin^{2}\theta,
\end{eqnarray}
where
\begin{eqnarray}
\bar{M} &=& 1-\frac{1}{2}\left(r-\sqrt{r^2+l^2}\right),
\\
\bar{\rho}^2 &=& r^2+l^2+a^2\cos^2\theta, \\
\bar{\Delta} &=& r^2+l^2+a^2-2\bar{M}\sqrt{r^2+l^2}, \\
\bar{A} &=& (r^2+l^2+a^2)^2-a^2 \bar{\Delta}\sin^{2}\theta.
\end{eqnarray}
Here $l$ is  the bounce radius. We select the time transformation
function  as
\begin{eqnarray}
 g=\frac{\bar{\rho}^2}{\bar{\Delta}}.
\end{eqnarray}
Obviously, $g\rightarrow 1$ for sufficiently large distance $r$.
Thus, the Hamiltonian (6) can be separated into explicitly
solvable several parts.

\section*{Acknowledgments}

The authors are very grateful to a referee for
valuable comments and suggestions. This research has been
supported by the National Natural Science Foundation of China
[Grant Nos. 11973020, 12073008 and 12473074].

\newpage

\begin{figure*}[ptb]
\center{
\includegraphics[scale=0.3]{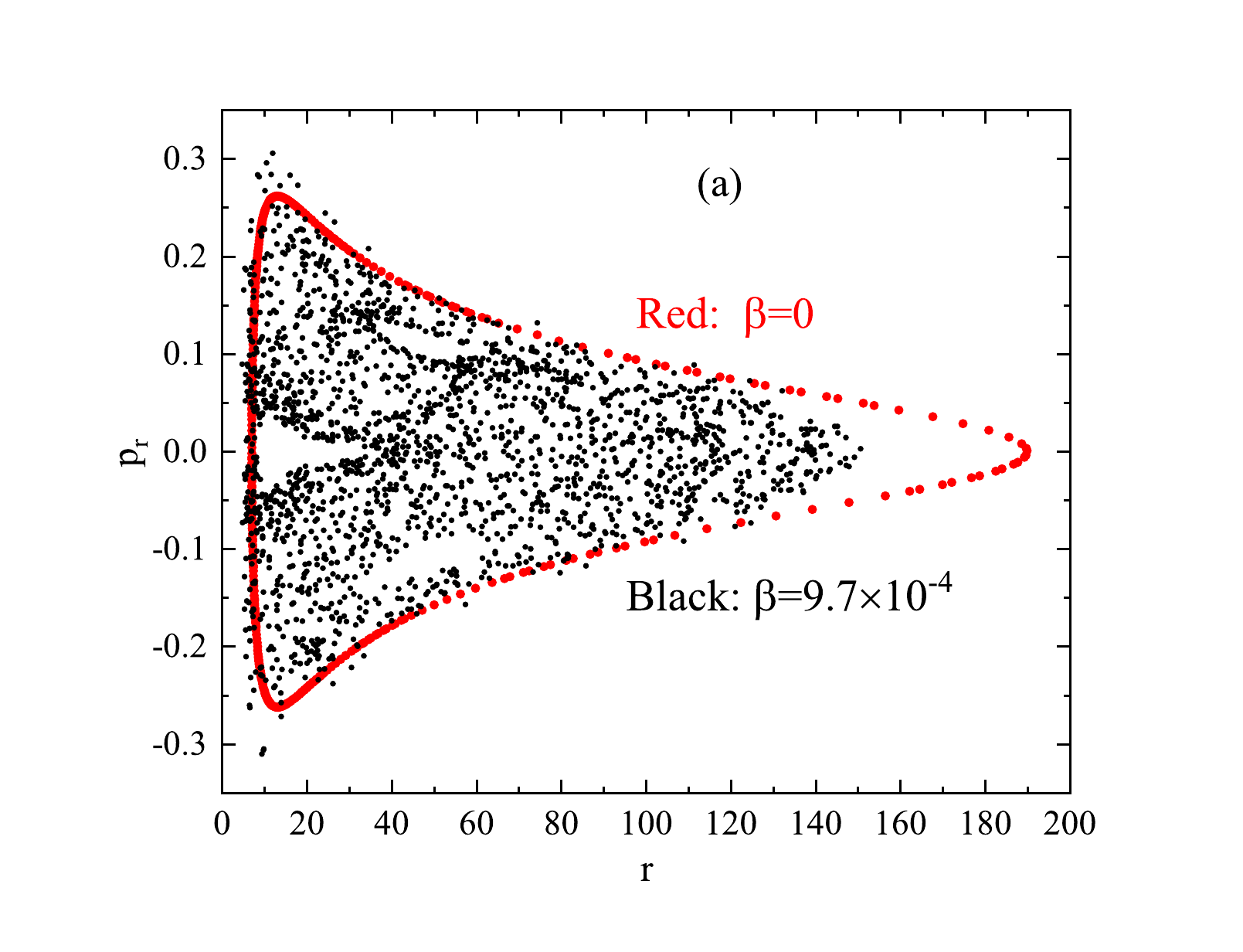}
\includegraphics[scale=0.3]{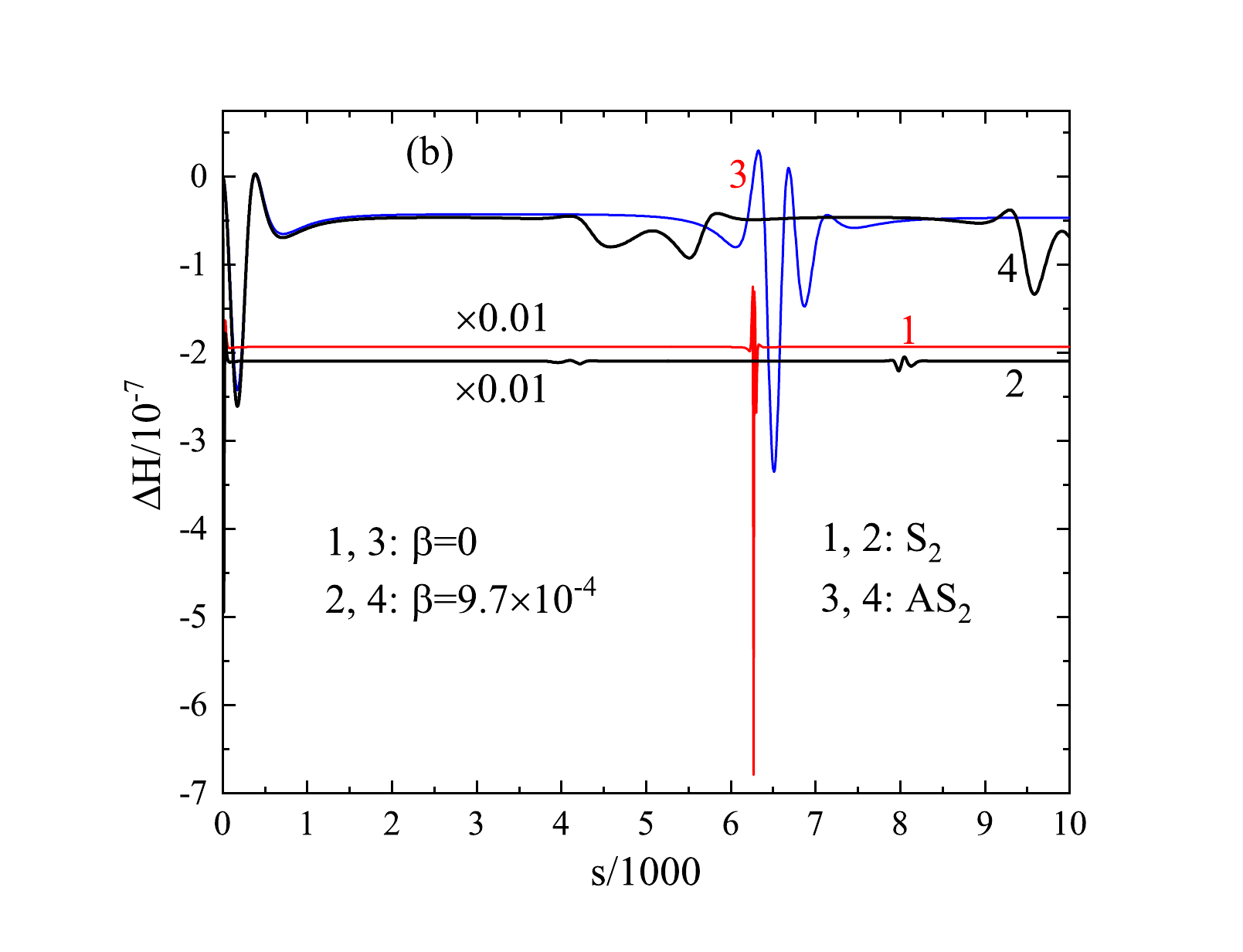}
\includegraphics[scale=0.3]{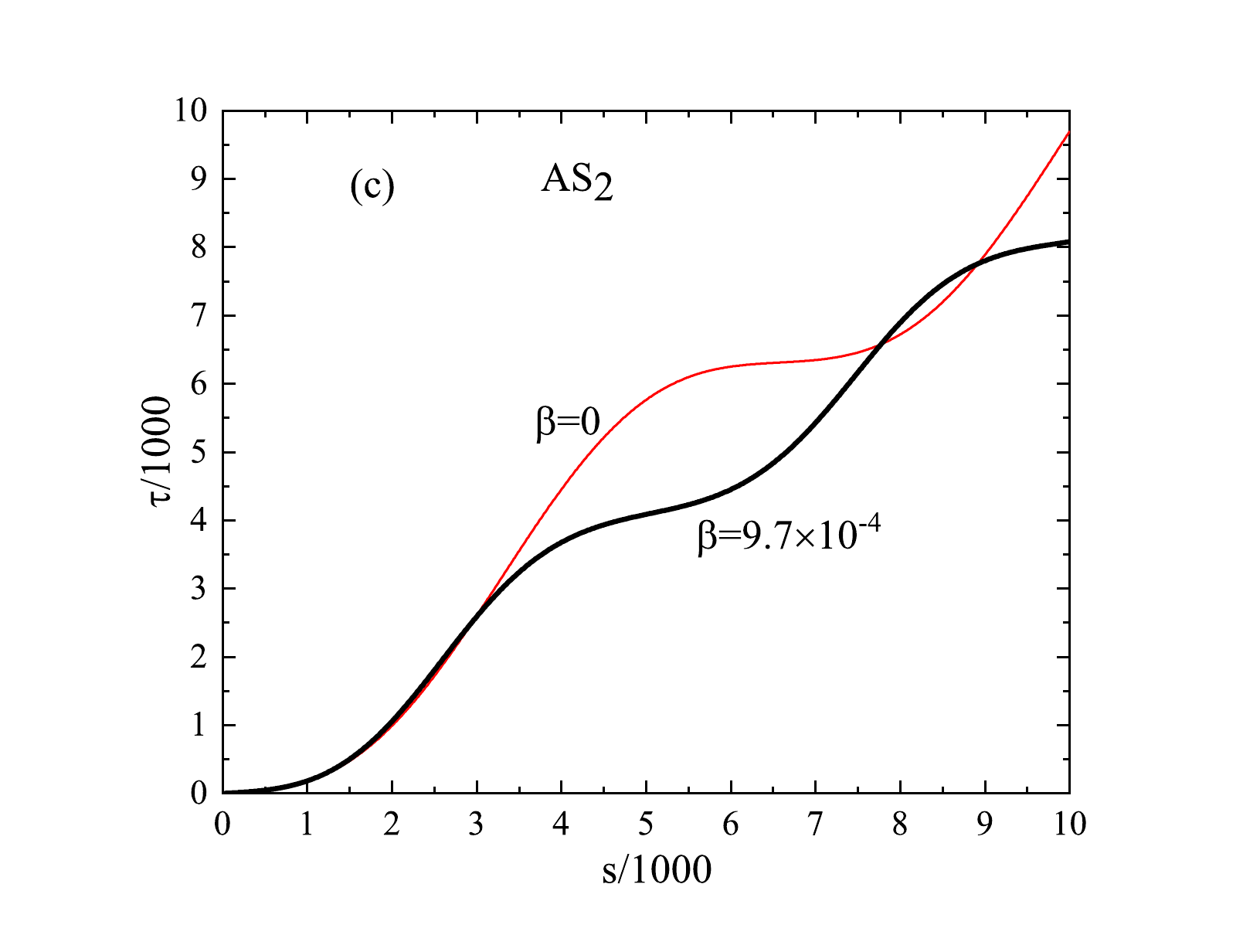}
\includegraphics[scale=0.3]{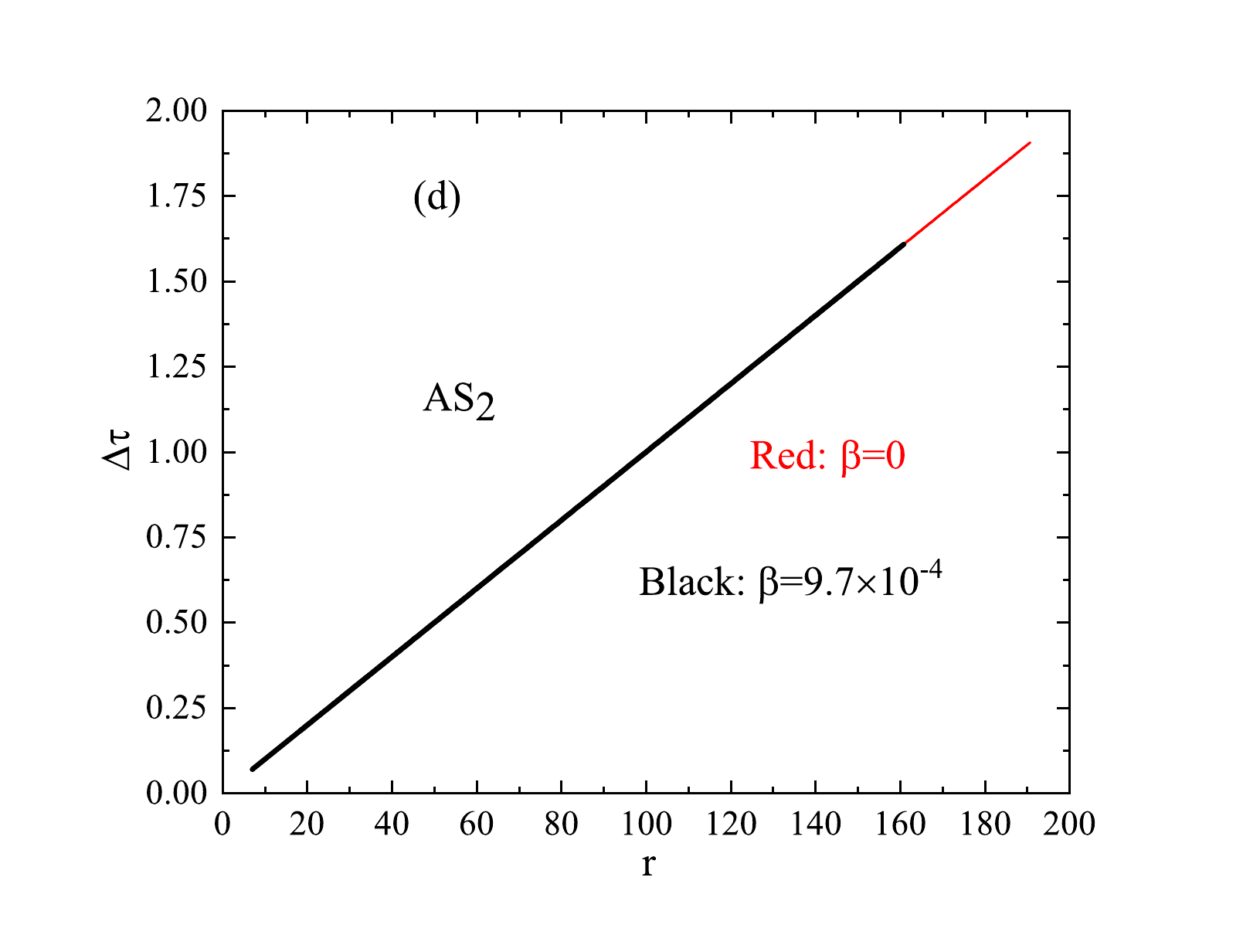}
\caption{Numerical comparisons in between two cases without
magnetic field $\beta=0$ and with magnetic field $\beta=9.7\times
10^{-4}$ in the Schwarzschild spacetime. (a) Phase space
structures of $r$ and $p_r$ described by Poincar\'{e} section at
the plane $\theta=\pi/2$.  The two algorithms $S_2$ and $AS_2$
almost give the same structures. (b) Hamiltonian errors $\Delta
H=H+p_0$. $''\times 0.01''$ for $S_2$ means that the plotted
errors are the practical ones multiplied by factor 0.01, and
$''s/1000''$ represents that the plotted new time $s$ is the
practical one divided by factor 1000. $s$ is $\tau$ for $S_2$. (c)
Relation between the original time $\tau$ and the new time $s$.
(d) Different proper time steps $\Delta\tau$ in different radial
separations $r$ of the orbit for $\beta=0$ or $\beta=9.7\times
10^{-4}$.} \label{Fig1} }
\end{figure*}

\begin{table*}[htbp]
\centering \caption{Effects of different choices of $j$ on the
errors $\Delta H$ and the proper times $\tau$. The orbit with
$\beta=0$ in Figure 1 is integrated by the method $AS_2$ until the
new time $s$ reaches 10,000. Only the orders of $\Delta H$ are
listed. $\tau$ is the proper time at the new time $s=10,000$. }
\label{Tab1}
\begin{tabular}{lcccccccccccc}
\hline $j$          & 1     & 10    &  50   & 100  & 200 & 1000  \\
\hline $\Delta H$   & $10^{-3}$ & $10^{-5}$ & $10^{-6}$  & $10^{-7}$  & $10^{-8}$  & $10^{-9}$  \\
\hline $\tau$       & $8.6\times10^{5}$ & $9.5\times10^{4}$ &
$1.9\times10^{4}$ & $9.7\times10^{3}$  & $5.7\times10^{3}$ & 178  \\
\hline
\end{tabular}
\end{table*}

\begin{figure*}[ptb]
\center{
\includegraphics[scale=0.2]{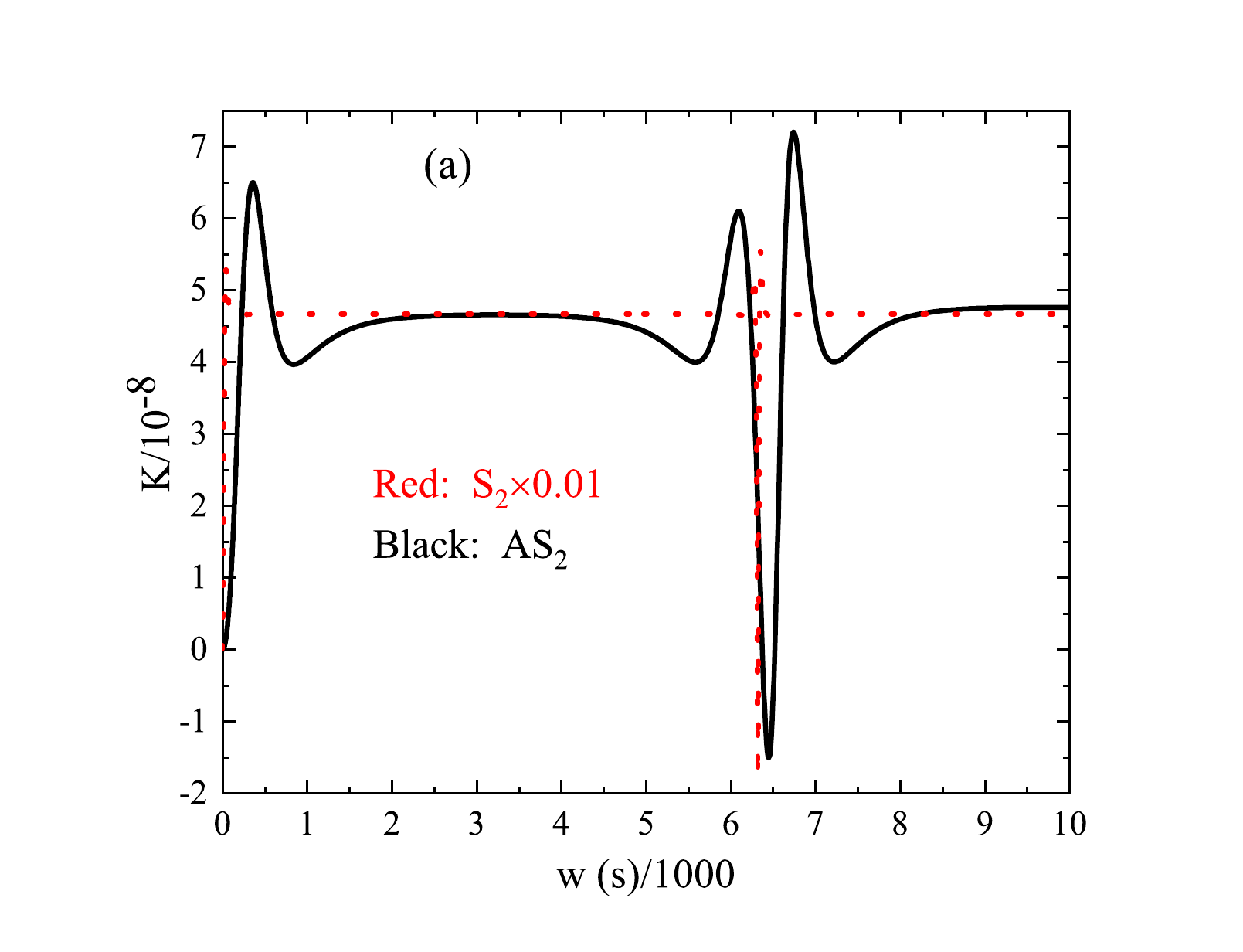}
\includegraphics[scale=0.2]{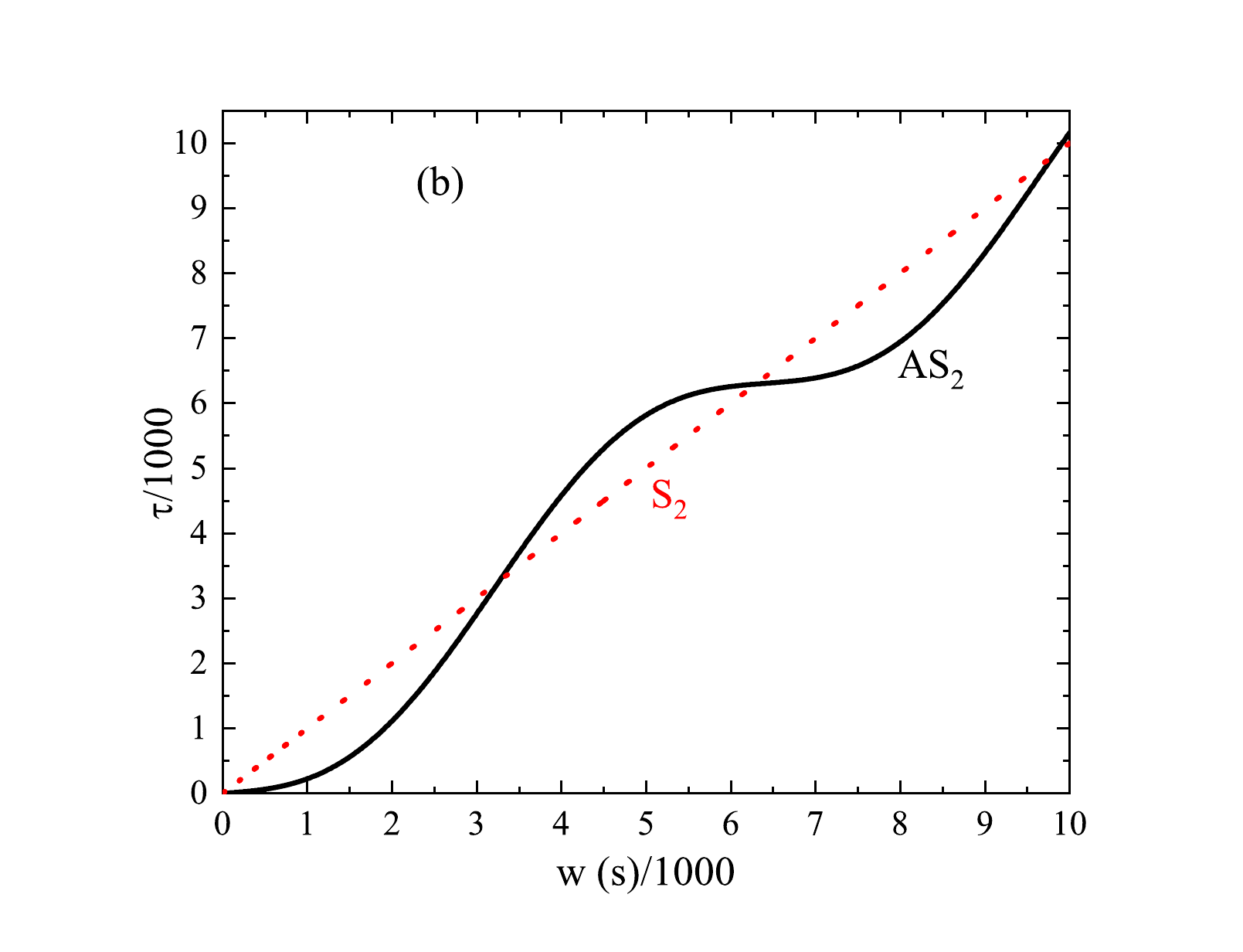}
\includegraphics[scale=0.2]{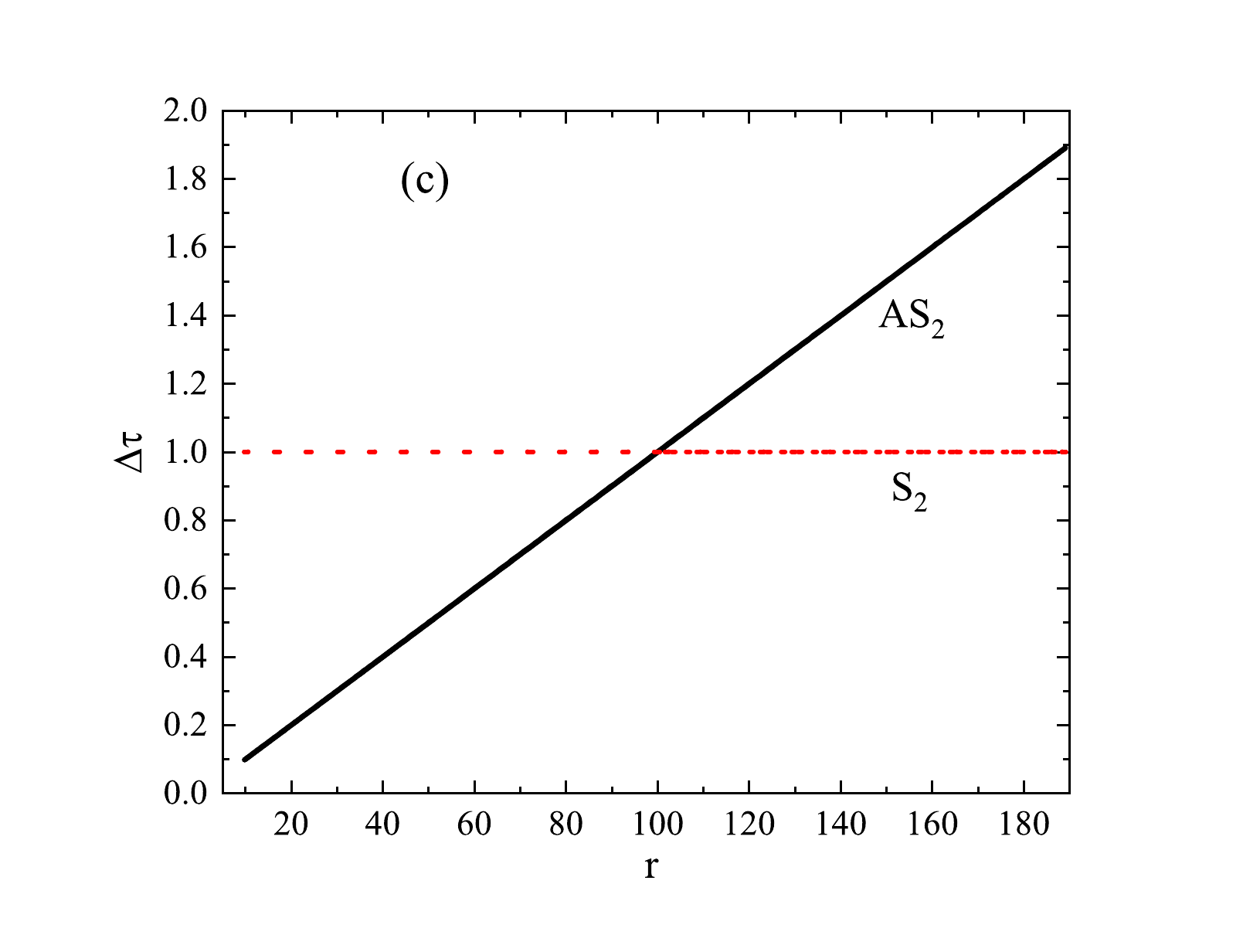}
\caption{Numerical comparisons in between the two methods $S_2$
and $AS_2$ integrating the time-like geodesics in the Kerr
spacetime. (a) Hamiltonian errors $K$ of Equation (6). The time is
$w$ for $S_2$, and $s$ for $AS_2$. The plotted errors are
decreased 100 times for $S_2$. (b) Relation between the proper
time $\tau$ and the new time $w$ or $s$. (c) Different proper time
steps $\Delta\tau$ in different radial separations $r$ of the
orbit.} \label{Fig2} }
\end{figure*}

\begin{figure*}[ptb]
\center{
\includegraphics[scale=0.3]{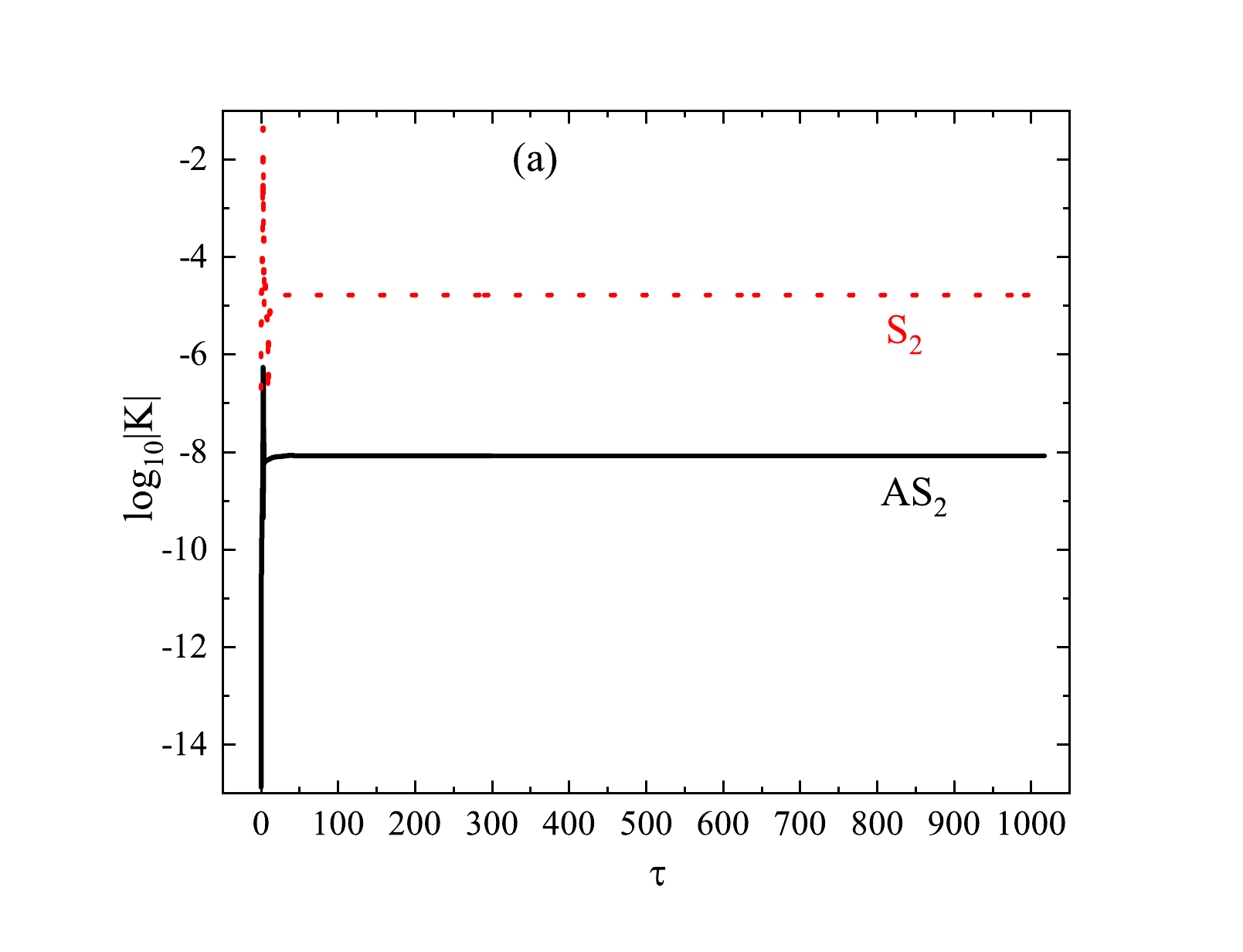}
\includegraphics[scale=0.3]{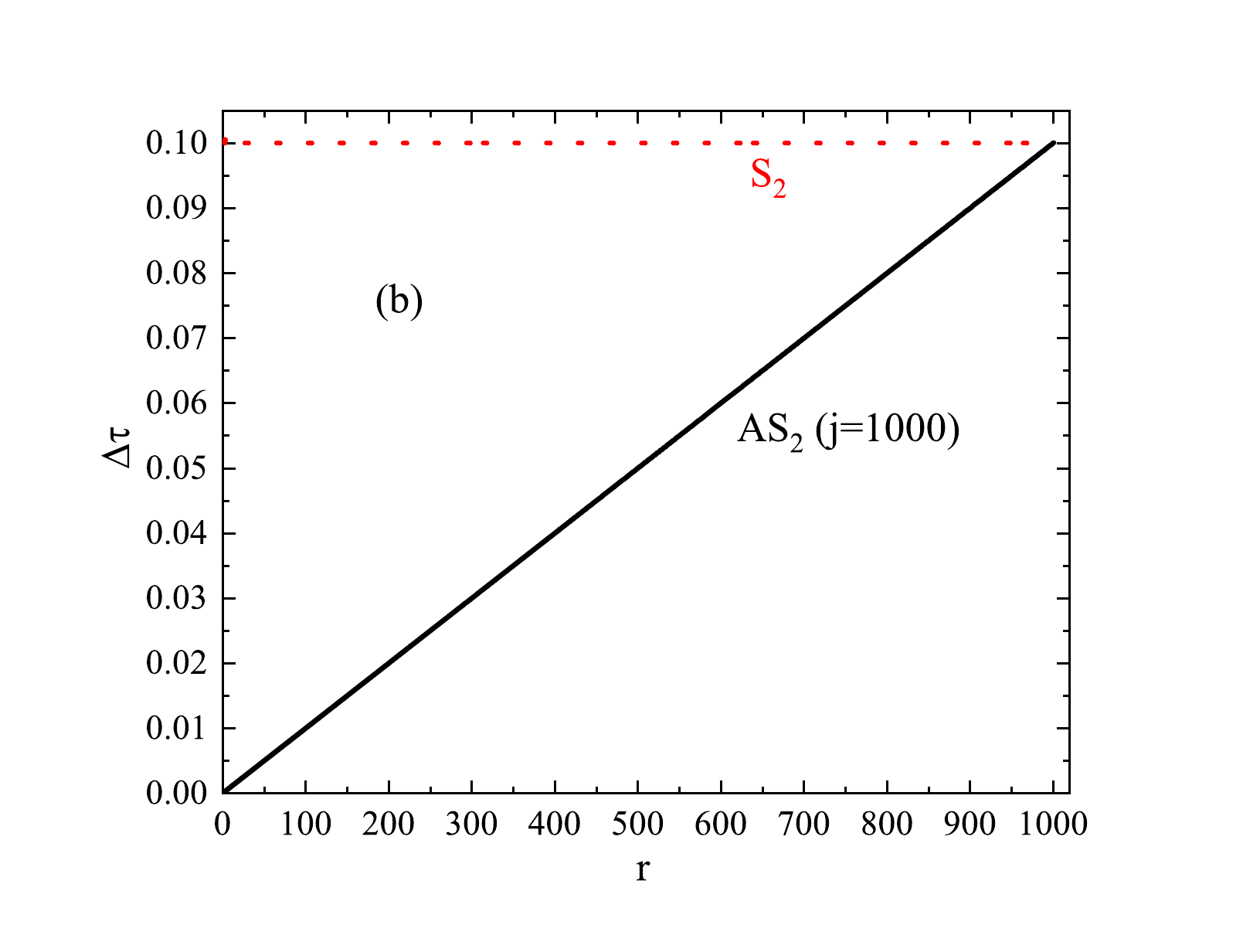}
\caption{Numerical comparisons in between the two methods $S_2$
and $AS_2$ integrating the null geodesics in the Kerr spacetime. A
photon orbit escaping to infinity has the parameters $a=0.3$,
$E=0.99$, $L=3\sqrt{3}E/10$ and the initial conditions $r=3$,
$\theta=\pi/2$, $p_r=0$. (a) Hamiltonian errors $K$ depending on
the old time $\tau$. (b) Different old time steps $\Delta\tau$ in
different radial separations $r$ of the escaping orbit.}
\label{Fig3} }
\end{figure*}

\begin{table*}[htbp]
\centering \caption{Effects of different choices of $j$ on the
errors $K$ and the new times $s$. The escaping photon orbit of
Figure 3 is integrated by the method $AS_2$ until the distance $r$
reaches 1000. The old time is $\tau\approx 1,016$. Only the orders
of $K$ are listed.} \label{Tab2}
\begin{tabular}{lcccccccccccc}
\hline $j$   & 10         & 100       &  500         & 1000       & 2000     \\
\hline $K$   & $10^{-4}$  & $10^{-6}$ & $10^{-7.5}$  & $10^{-8}$  & $10^{-8.7}$    \\
\hline $s$   & 81         & 811       & 4,052        & 8,103      & 16,205     \\
\hline
\end{tabular}
\end{table*}

\begin{figure*}[ptb]
\center{
\includegraphics[scale=0.3]{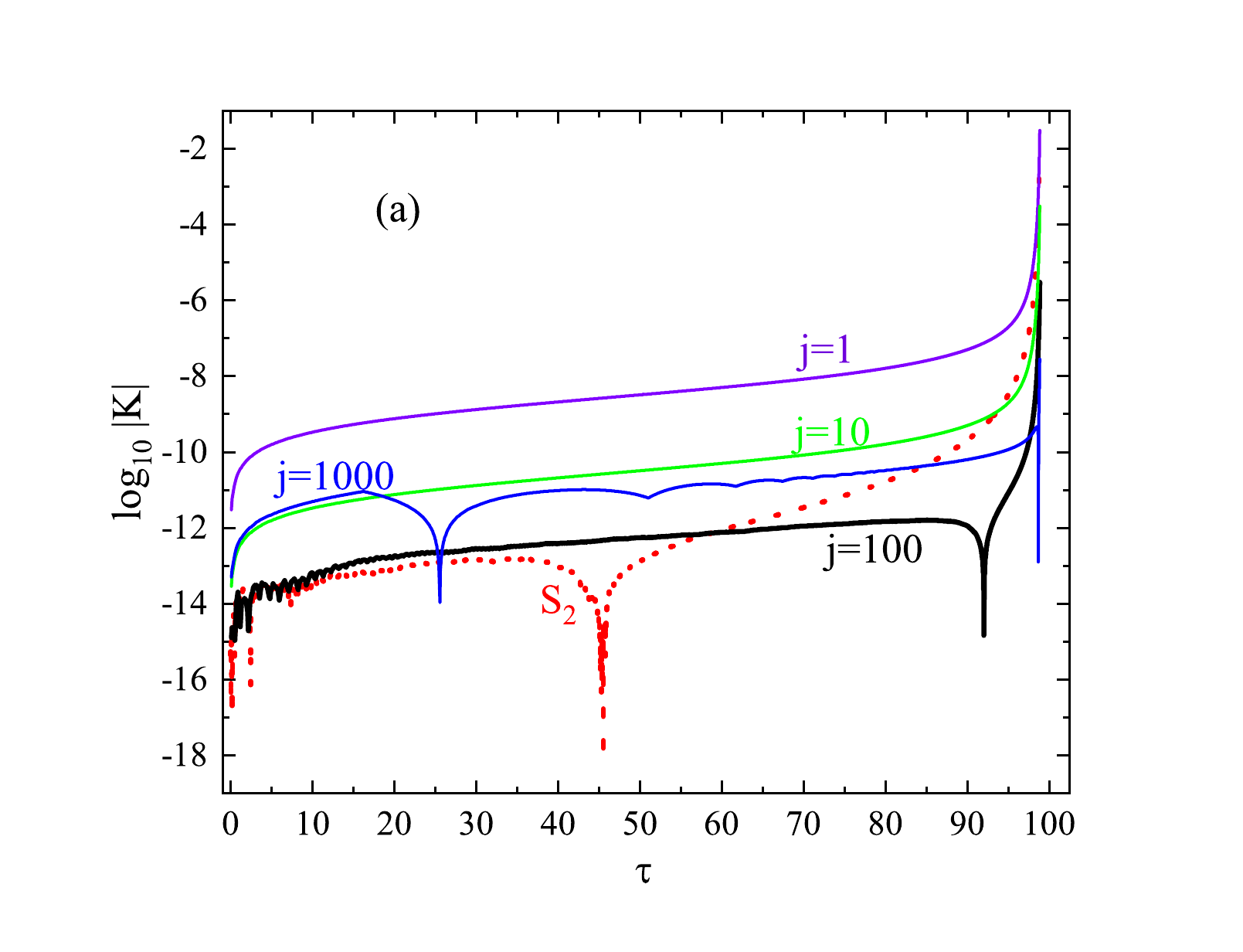}
\includegraphics[scale=0.3]{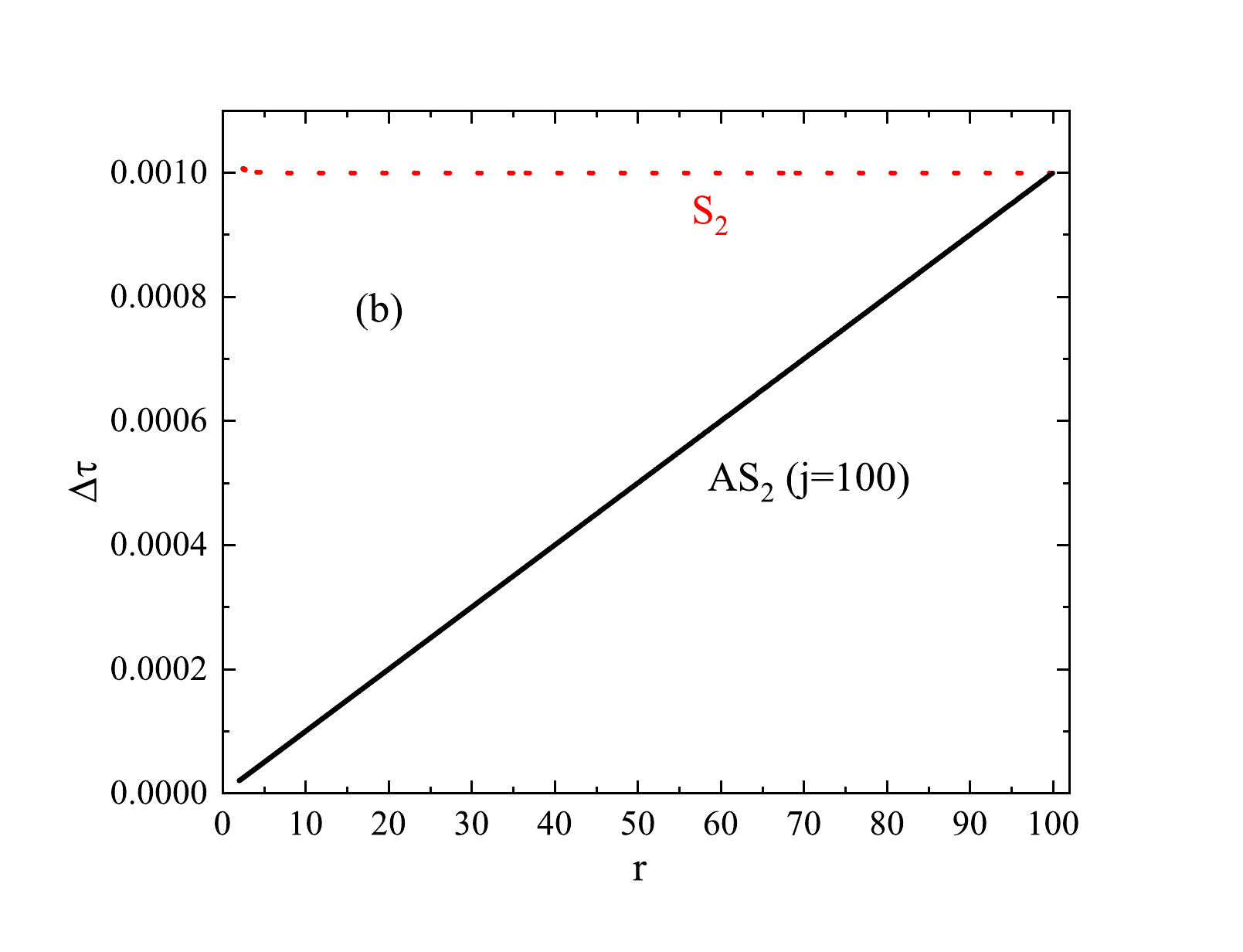}
\caption{Same as Figure 3, but a photon orbit falling into the
black hole with the parameters $a=0.5$, $E=0.996$, $L=-2.211$ and
the initial conditions $r=100$, $\theta=\pi/2$, $p_r=-1.016$ is
integrated. (a) Hamiltonian errors $K$; the best choice is
$j=100$. (b) Different old time steps $\Delta\tau$ in different
radial separations $r$ of the falling orbit. } \label{Fig4} }
\end{figure*}

\begin{figure*}[ptb]
\center{
\includegraphics[scale=0.3]{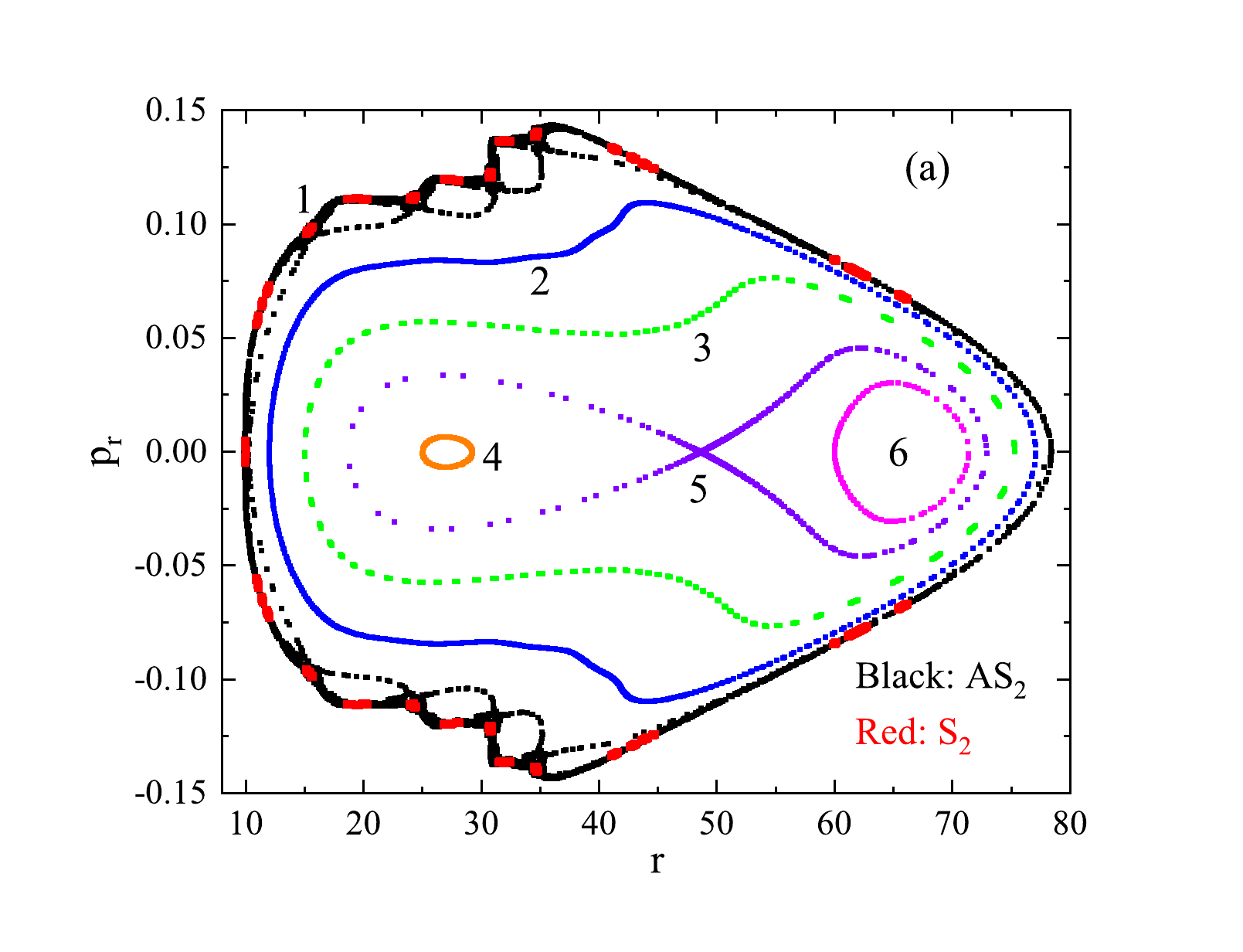}
\includegraphics[scale=0.3]{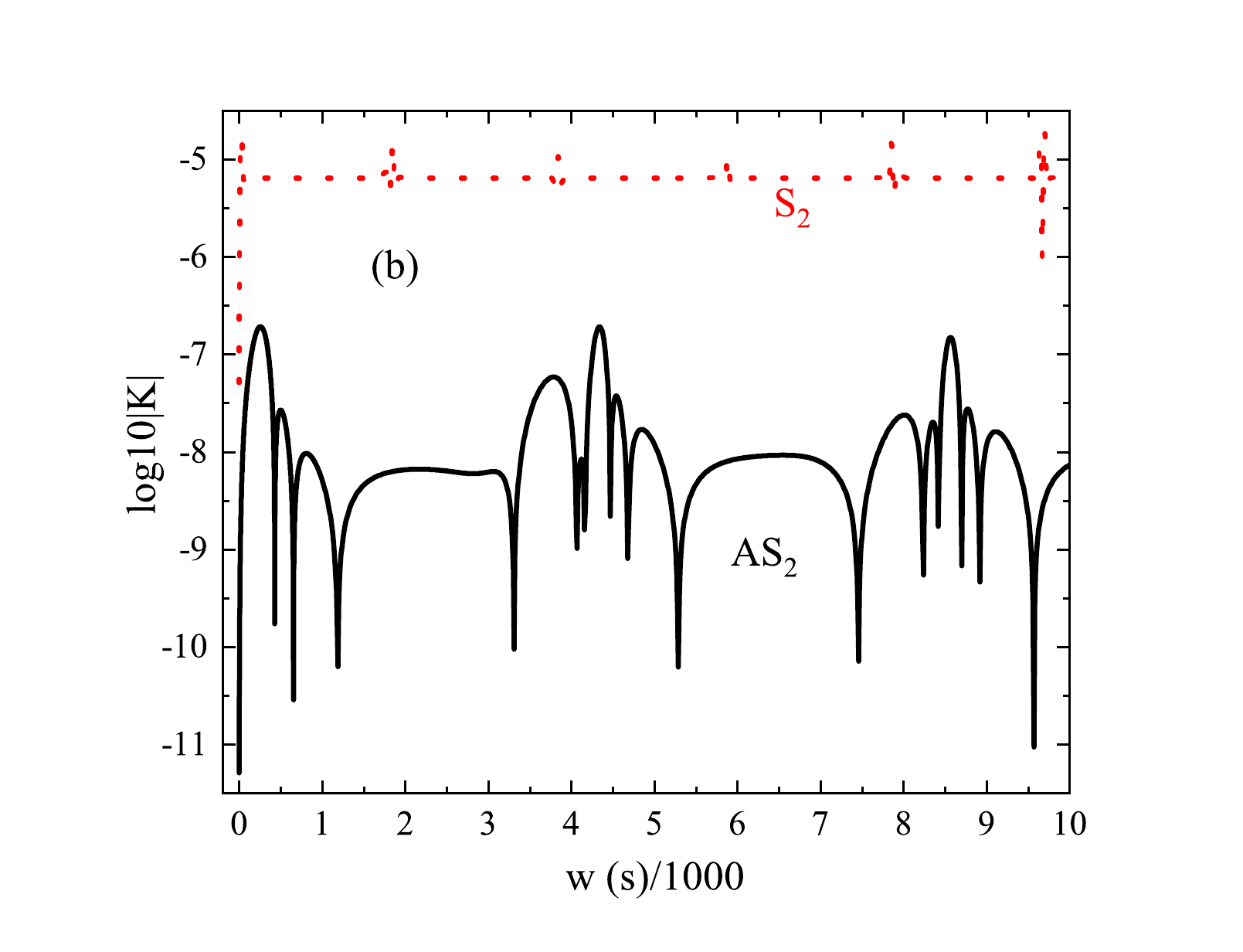}
\includegraphics[scale=0.3]{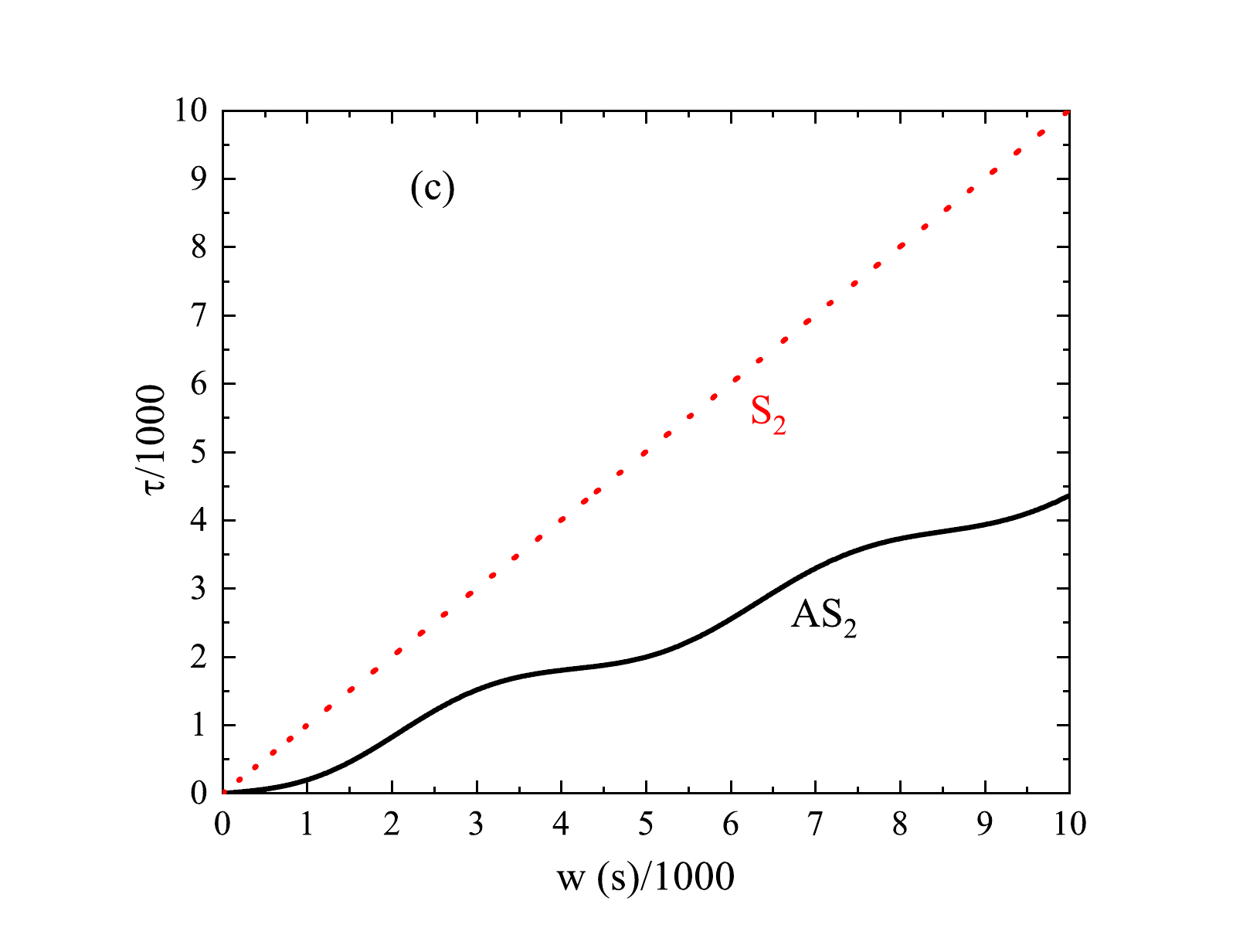}
\includegraphics[scale=0.3]{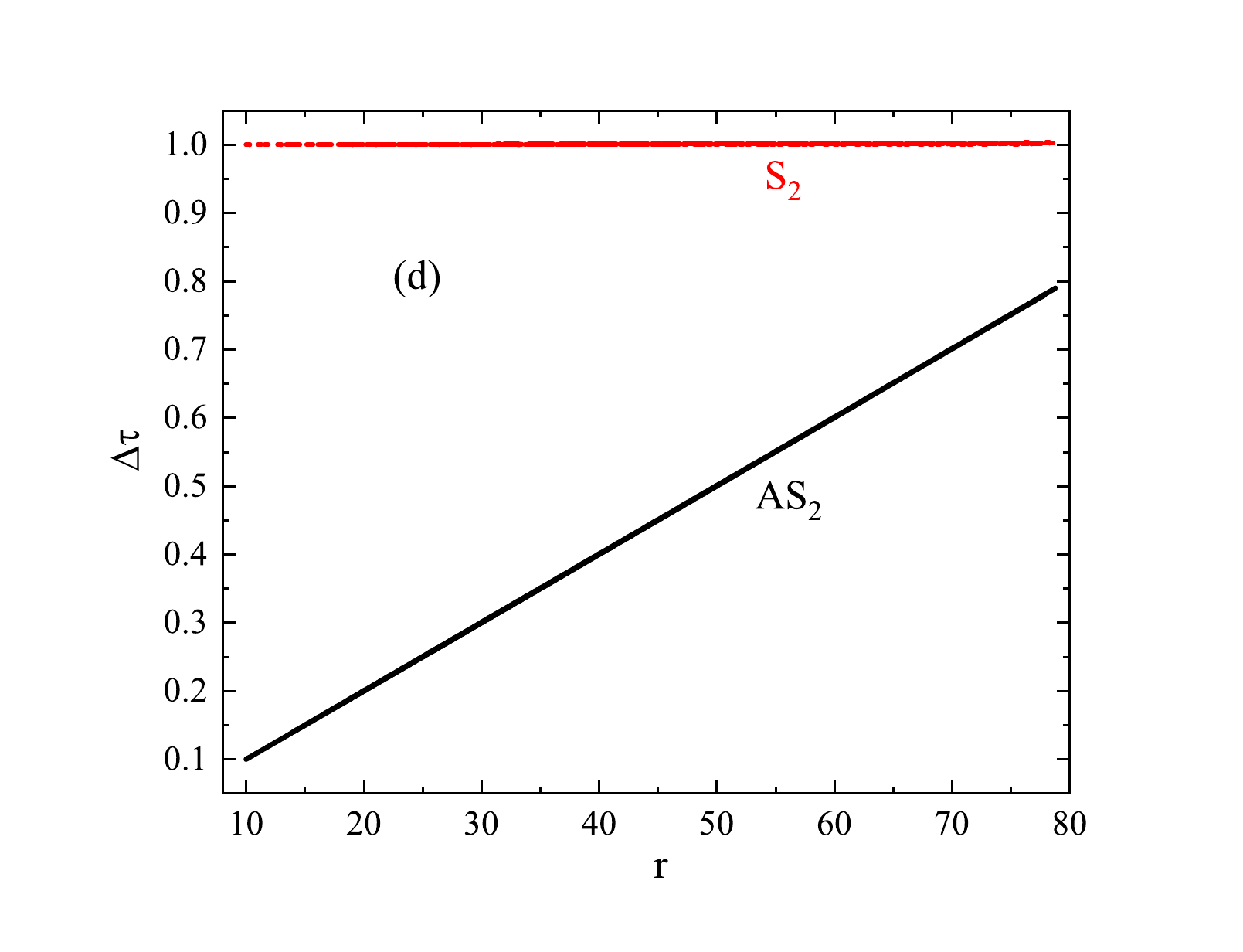}
\caption{Numerical comparisons in simulations of particles around
the Schwarzschild-Melvin black hole. The step size is $h=1$, and
the parameters are $E=0.9905$, $L=3.6$, $B=0.001$ and $j=100$. (a)
Poincar\'{e} section. Orbit 1 with the initial separation $r=10$
seems to be regular for $S_2$ (colored Red), but is weakly chaotic
for $AS_2$ (colored Black). Both methods almost give the same
results to any one of Orbits 2-6. (b) Accuracies of Hamiltonian
$K$ in integrations of Orbit 1. (c) Relation between $w$ (or $s$)
and $\tau$. (d) Relation between  the proper time step
$\Delta\tau$ and the distance $r$. } \label{Fig5} }
\end{figure*}

\begin{figure*}[ptb]
\center{
\includegraphics[scale=0.3]{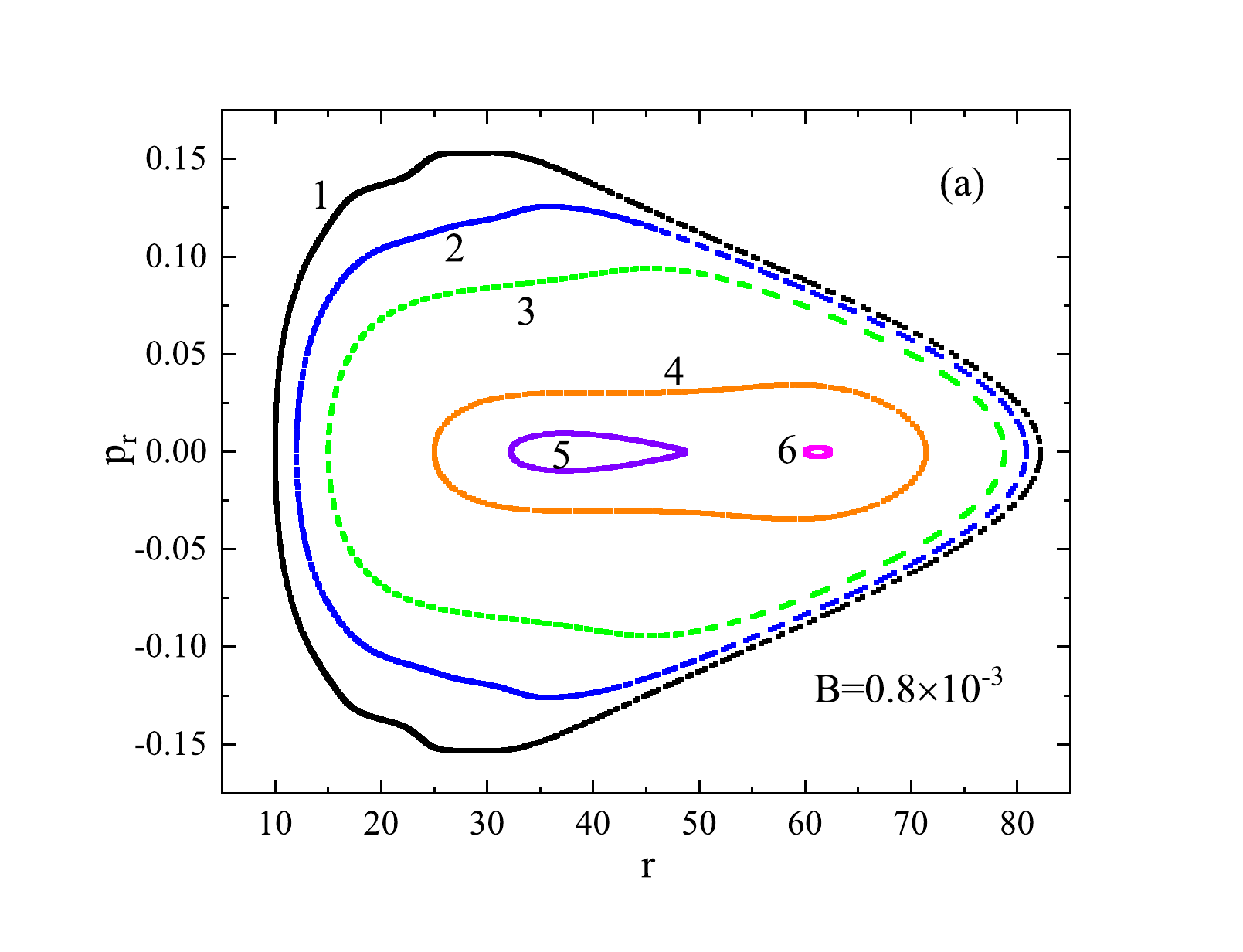}
\includegraphics[scale=0.3]{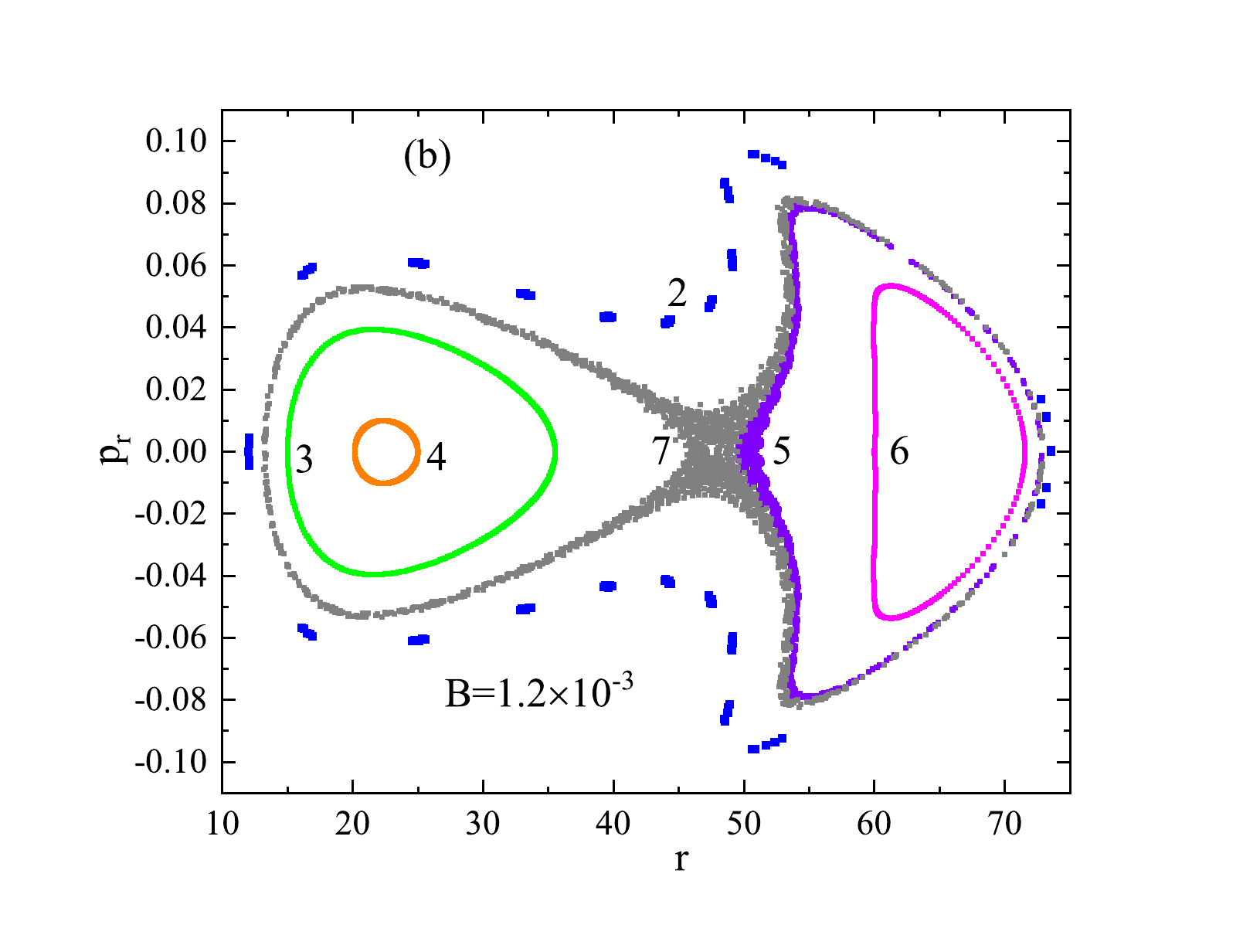}
\caption{Same as Figure 5a but only for the use of $AS_2$ in two
different values of magnetic field parameter $B$. (a) In the case
of $B=0.8\times 10^{-3}$, all orbits are regular KAM tori, and the
shapes of Orbits 2-6 are unlike those in Figure 5a. (b) When
$B=1.2\times 10^{-3}$, Orbit 5 with the initial distance $r=48.7$
and Orbit 7  with the initial distance $r=45$ are chaotic. }
\label{Fig6} }
\end{figure*}

\begin{figure*}[ptb]
\center{
\includegraphics[scale=0.2]{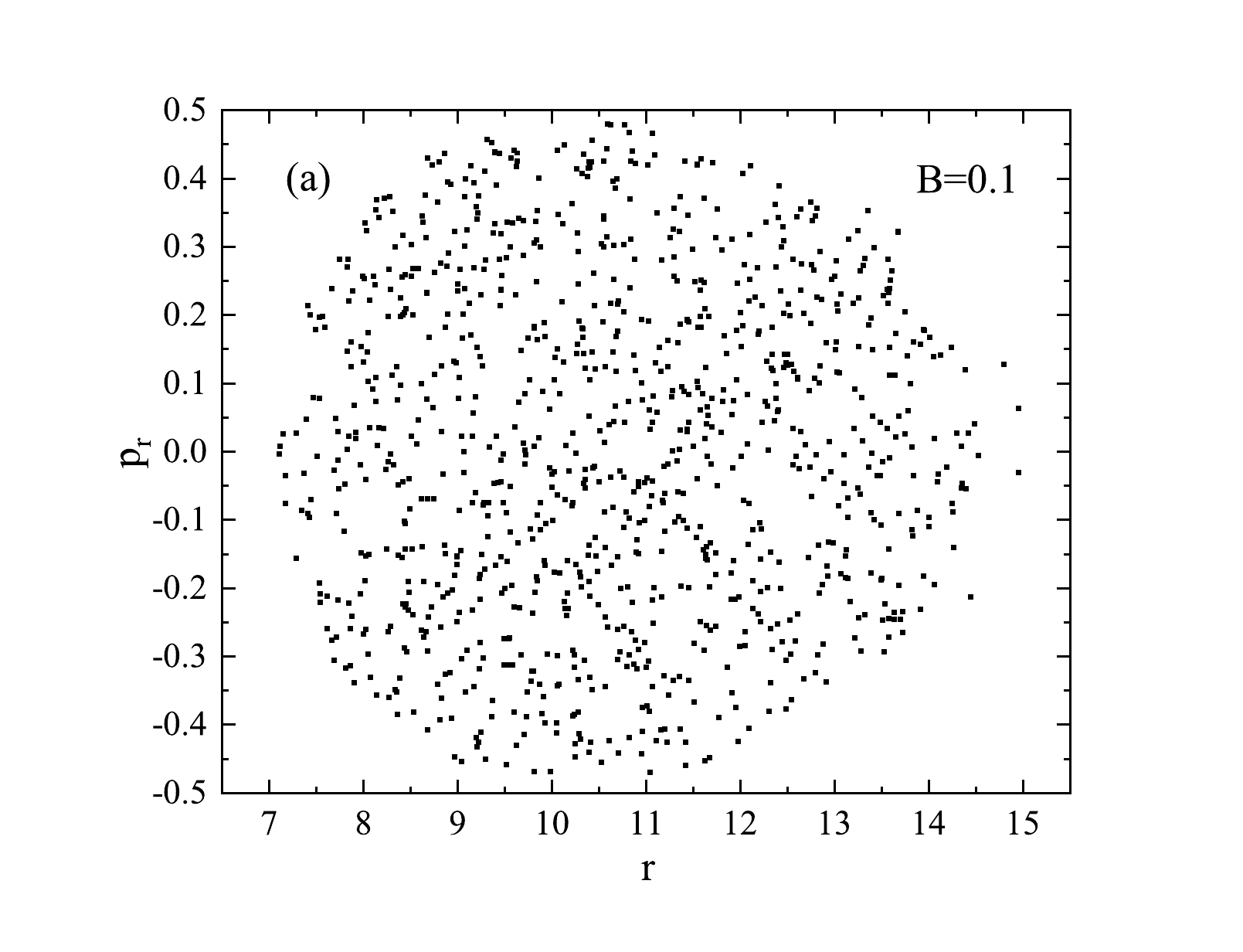}
\includegraphics[scale=0.2]{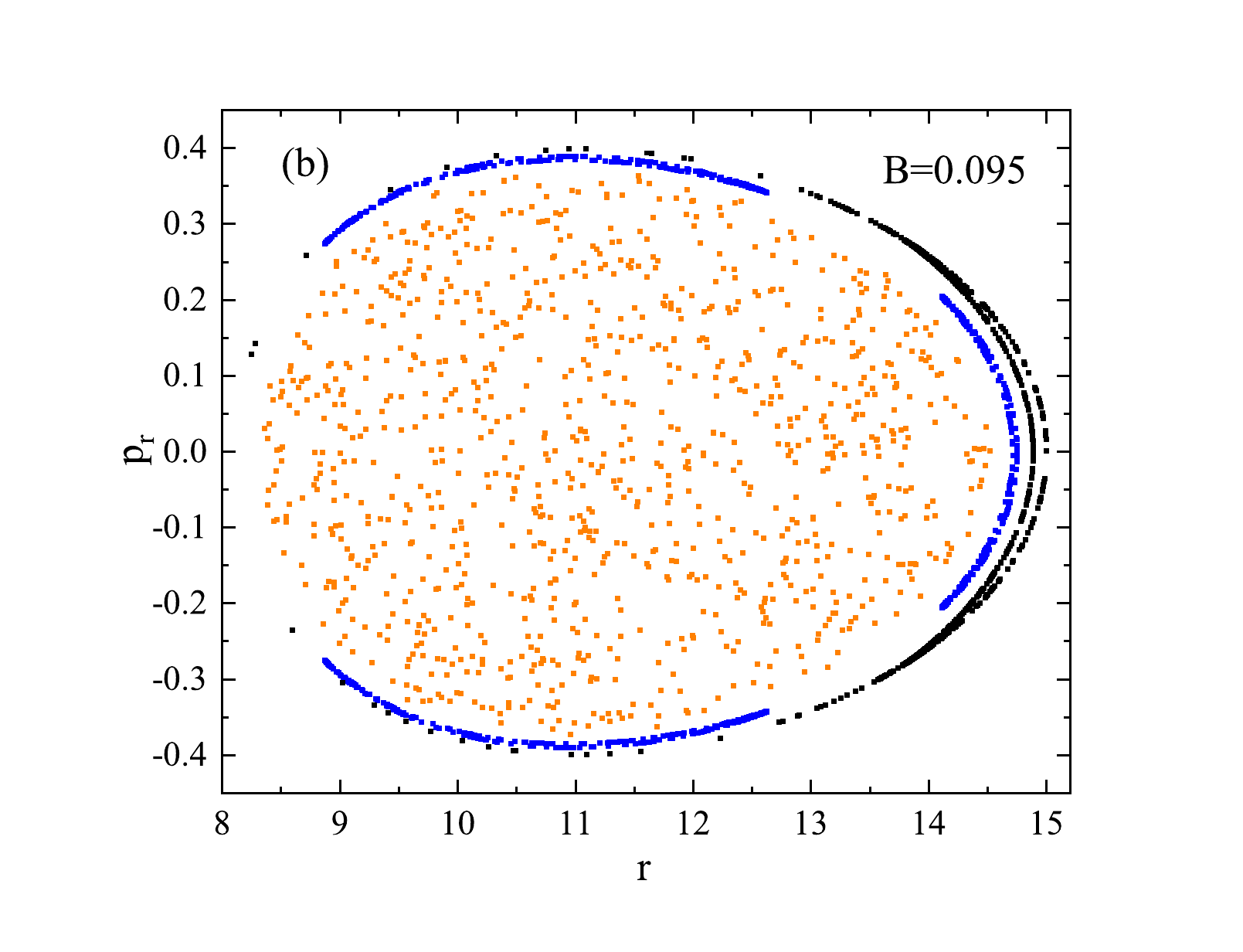}
\includegraphics[scale=0.2]{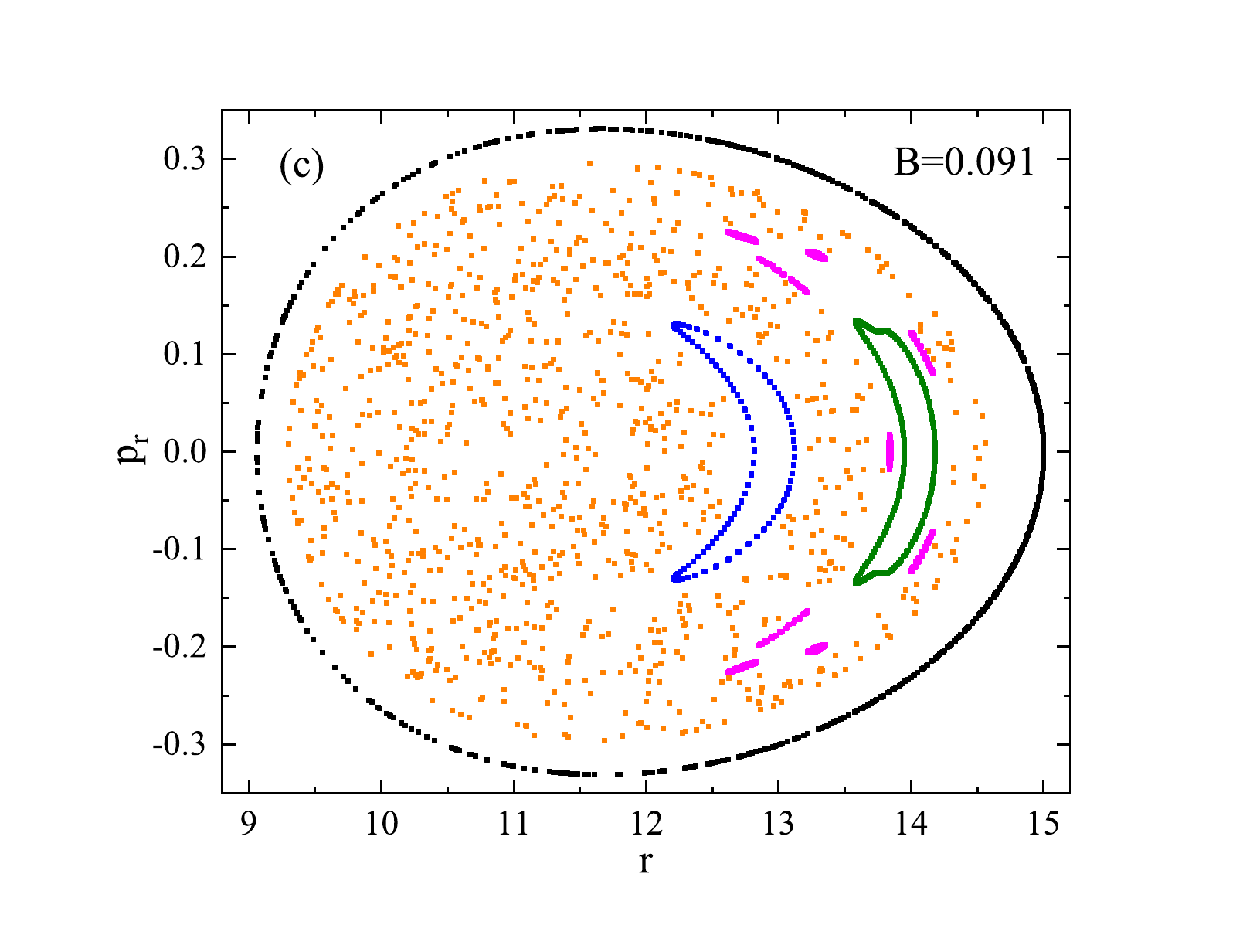}
\includegraphics[scale=0.2]{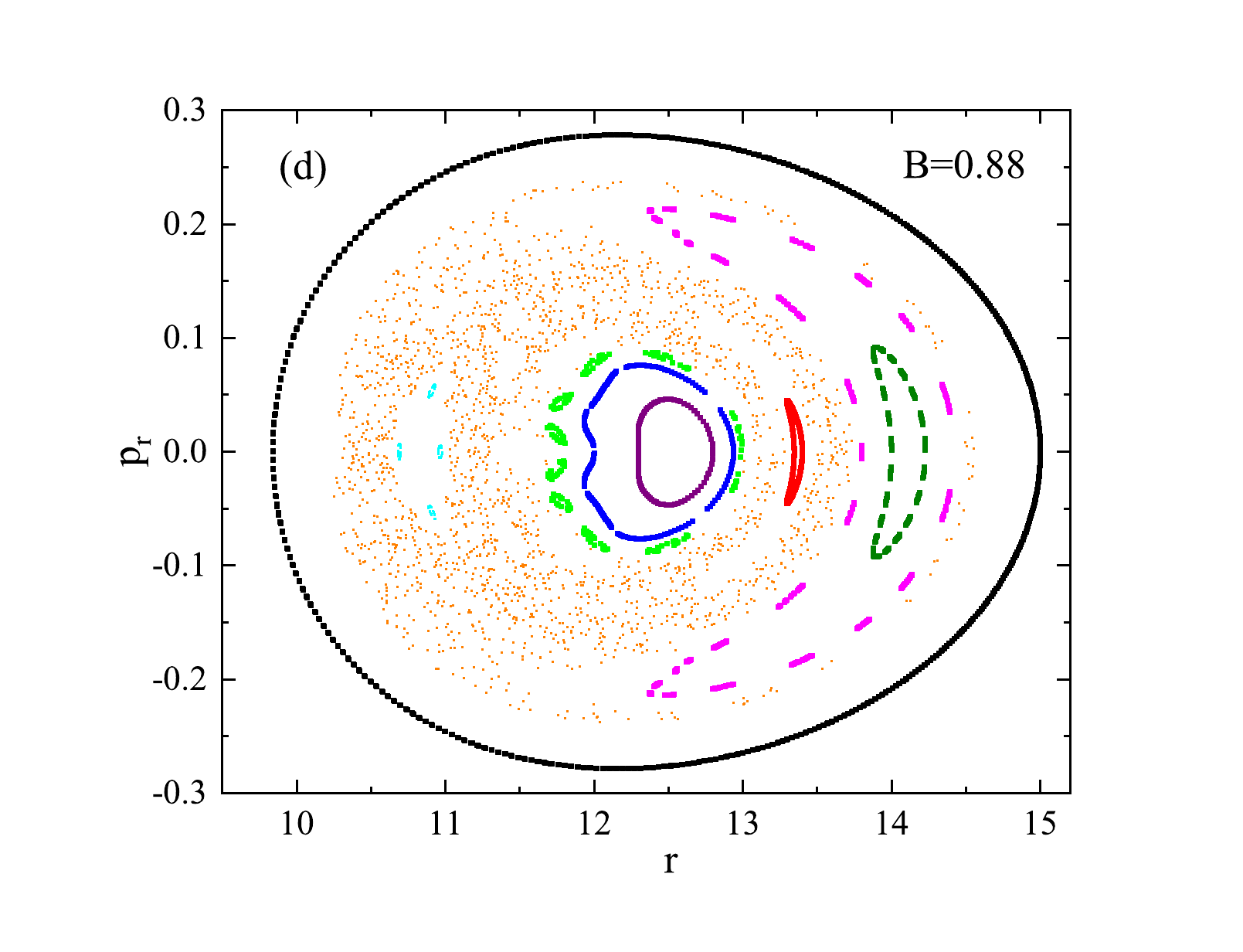}
\includegraphics[scale=0.2]{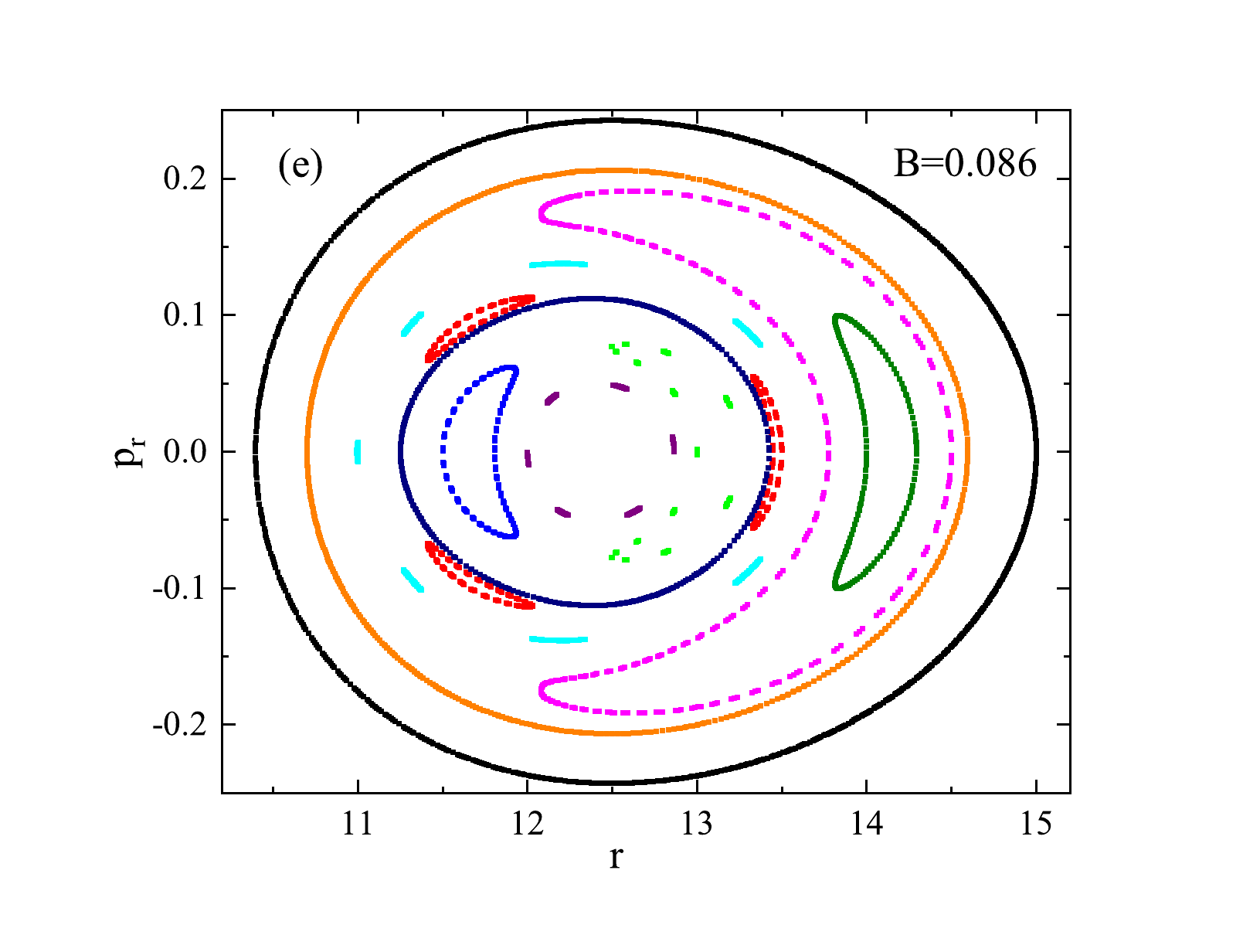}
\includegraphics[scale=0.2]{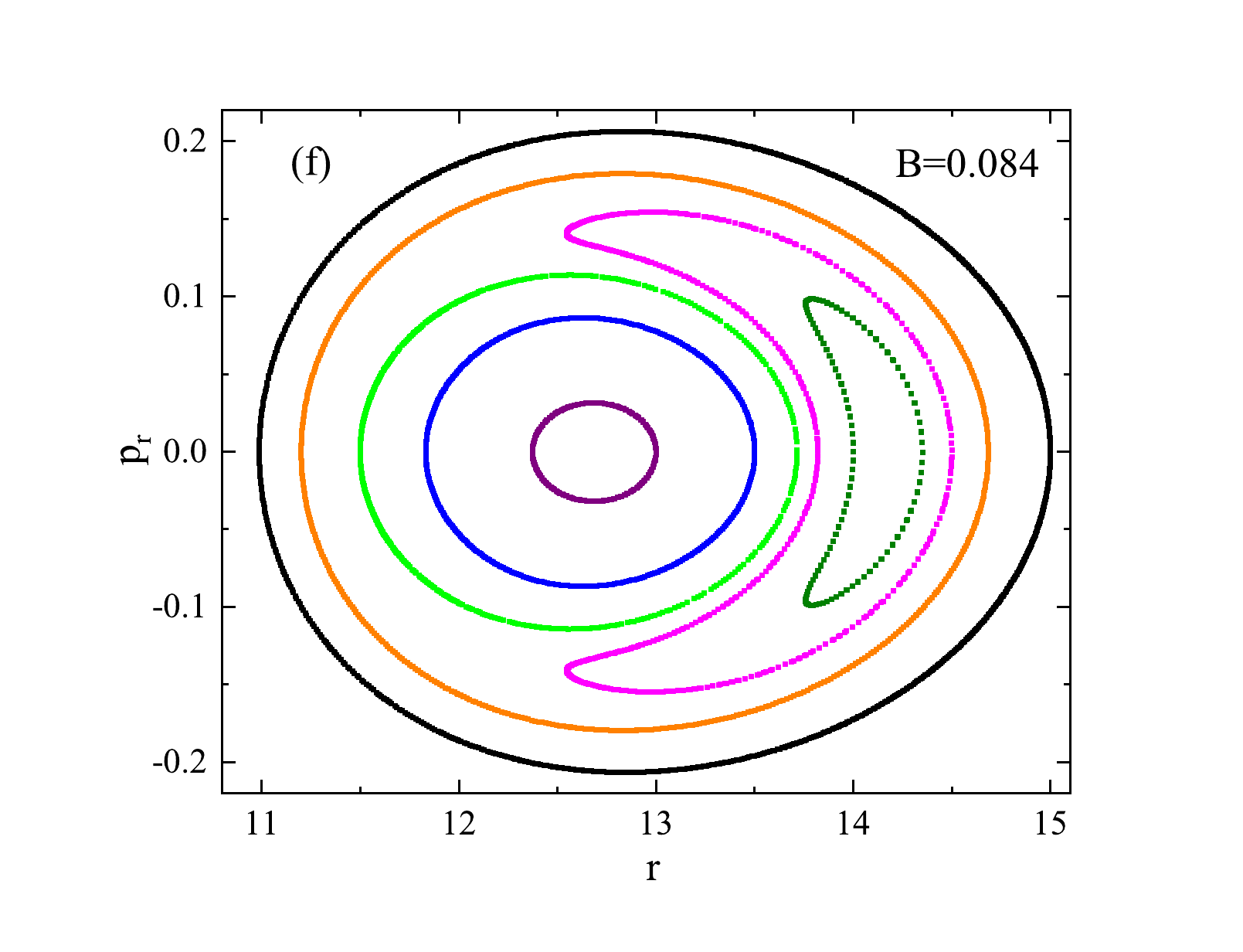}
\caption{Poincar\'{e} sections for the motions of photons near the
Schwarzschild-Melvin black holes with several values of the
magnetic field. The method $AS_2$ with the step size $h=1$ is
adopted. The other parameters are $E=0.995$ and $j=100$. The
angular momentum is always given by $L=-E\sqrt{-g_{\phi\phi}
/g_{tt}}$ with $r=15$. The extent of chaos is weakened as the the
magnetic field strength $B$ decreases in panels (a)-(f). }
\label{Fig7} }
\end{figure*}

\begin{figure*}[ptb]
\center{
\includegraphics[scale=0.2]{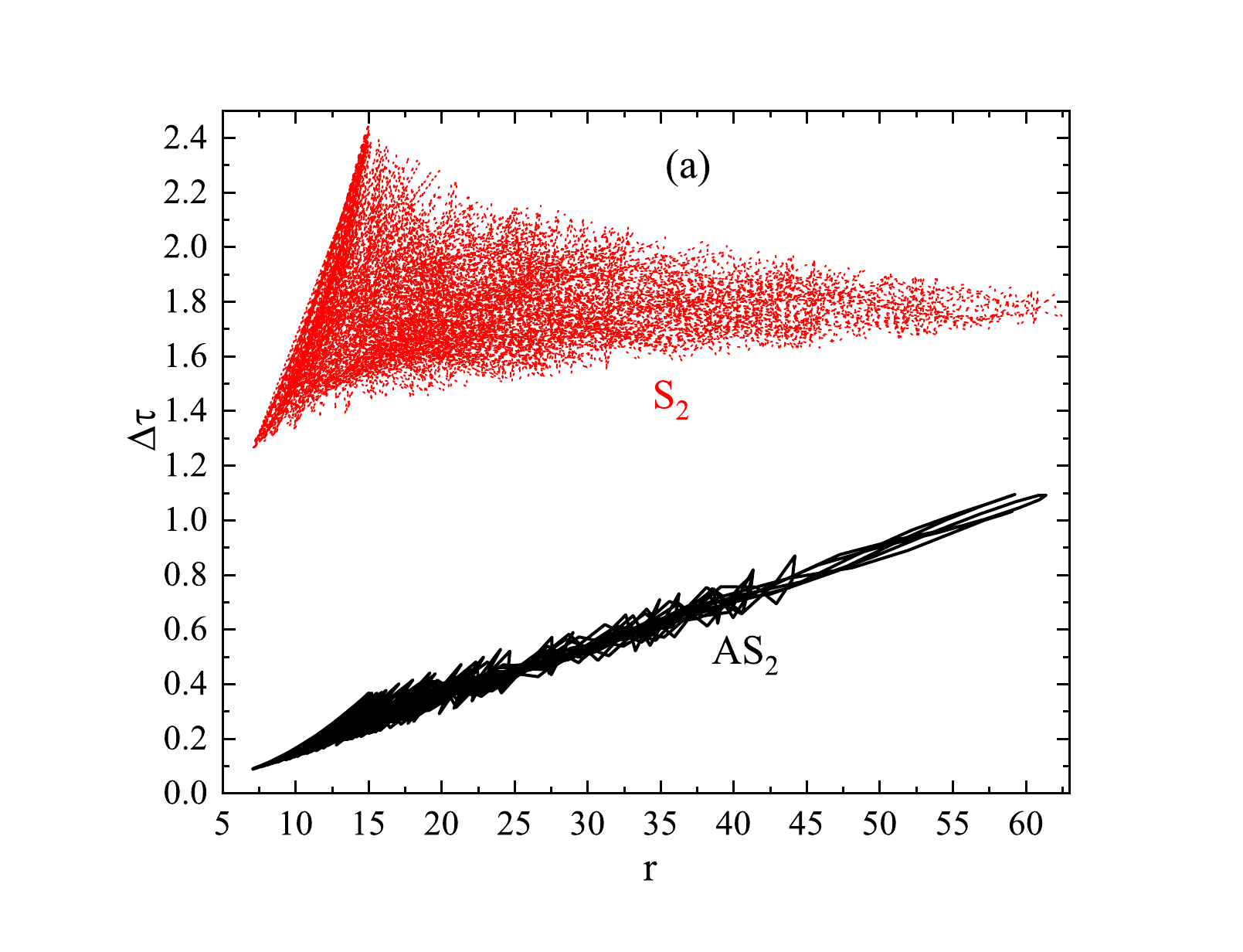}
\includegraphics[scale=0.2]{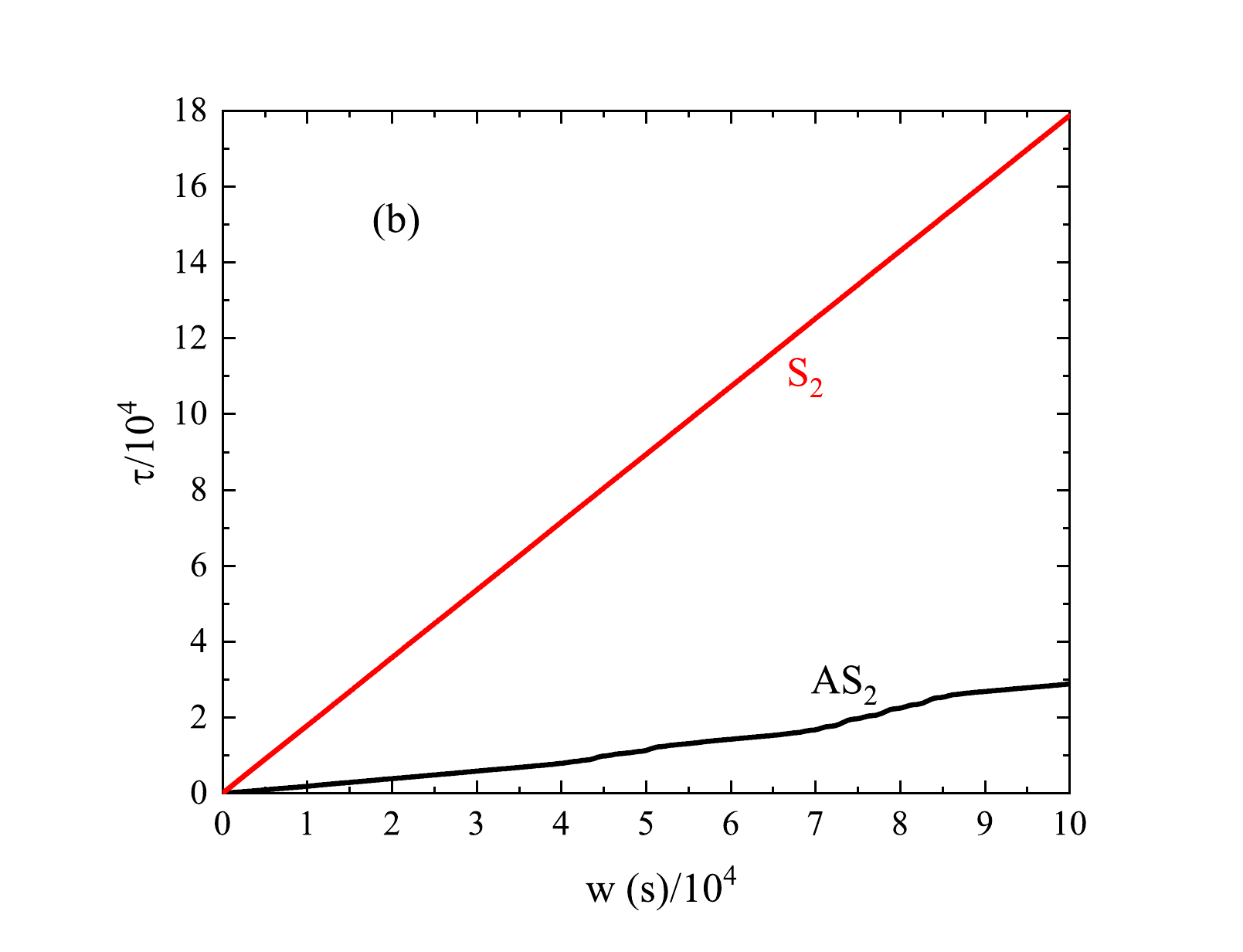}
\includegraphics[scale=0.2]{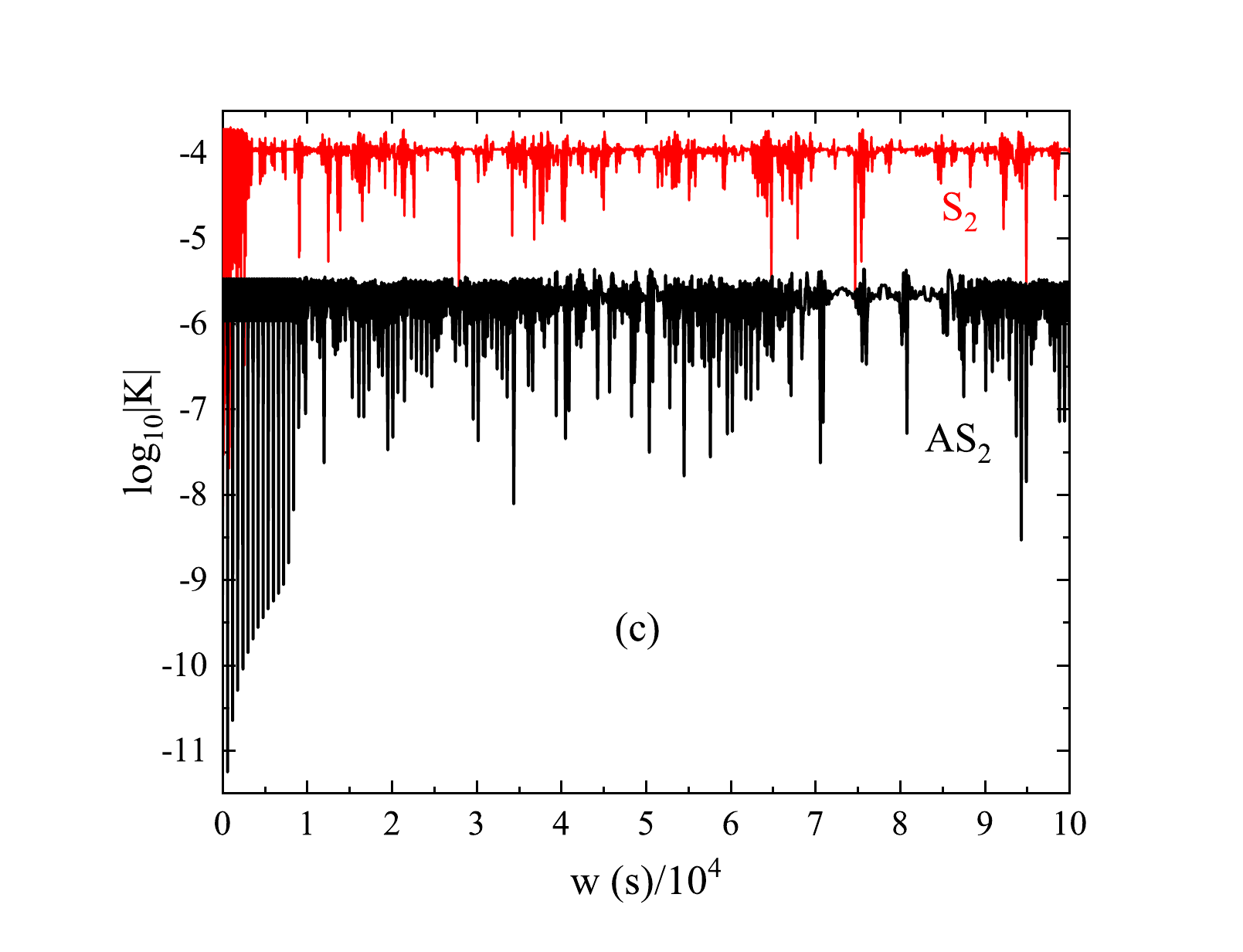}
\caption{Numerical comparisons between the methods $S_2$ and
$AS_2$ integrating the chaotic photon orbit with the initial
separation $r=15$ of Figure 7a. The integration step number is
$10^5$. (a) The old time steps $\Delta\tau$ depending on the
distance $r$. (b) The original time $\tau$ depending on the
transformation time $w$ or $s$. (c) Accuracy of Hamiltonian $K$. }
\label{Fig8} }
\end{figure*}

\end{document}